\documentclass[twoside,draft,reqno,12pt]{amsart}
\usepackage[english]{babel}
\usepackage{amsmath}
\usepackage{amssymb,ulem}
\usepackage{enumerate}

\usepackage{amsfonts}
\usepackage{graphicx}

\usepackage[ citecolor=black, linkcolor=black,
colorlinks, hypertexnames=false, plainpages=false]{hyperref}

\usepackage{color}
\newtheorem{lemma}     {Lemma}[section]
\newtheorem{thm}   [lemma]{Theorem}
\newtheorem{teorema1}   [lemma]{Theorem}

\newtheorem{coro}       [lemma]{Corollary}
\newtheorem{cong1}      [lemma]{Conjecture}
\newtheorem{remark1}    [lemma]{Remark}
\newtheorem{defin}      [lemma]{Definition}

\numberwithin{equation}{section}

\newcommand{\dis}{\displaystyle}
\newcommand{\e}{\epsilon}
\newcommand{\arctanh}{\mathrm{arctanh}\,}

\newcommand{\card}{\mathrm{card}\,}

\def\argmax{\mathrm{argmax}}
\def\argmin{\mathrm{argmin}}

\begin{document}

\title[Large deviations for the macroscopic motion of an interface]
{Large deviations for the macroscopic motion of an interface}

\author{P. Birmpa}
\address{Panagiota Birmpa,
Department of Mathematics, University of Sussex \newline
\indent Brighton, U.K.}
\email{P.Birmpa@sussex.ac.uk}

\author{N. Dirr}
\address{Nicolas Dirr,
Department of mathematics, Cardiff University \newline
\indent Cardiff, U.K.}
\email{DirrNP@cardiff.ac.uk}

\author{D. Tsagkarogiannis}
\address{Dimitrios Tsagkarogiannis,
Department of Mathematics, University of Sussex \newline
\indent Brighton, U.K.}
\email{D.Tsagkarogiannis@sussex.ac.uk}

\begin{abstract}
We study the most probable way an
interface moves on a macroscopic scale from an initial
to a final position within a fixed time in the context of large deviations for a stochastic microscopic lattice system of Ising spins with Kac interaction evolving in time according to Glauber (non-conservative) dynamics.
Such interfaces separate two stable phases of a ferromagnetic system and
in the macroscopic scale are represented by sharp transitions.
We derive quantitative estimates for the upper and the lower bound of
the cost functional that penalizes all possible deviations and obtain explicit
error terms which are valid also in the macroscopic scale.
Furthermore, using the result of a companion paper about the minimizers of
this cost functional for the macroscopic motion of the interface in a fixed time, 
we prove that the probability of such events can concentrate
on nucleations should the transition happen fast enough.
\end{abstract}


\maketitle

\section{Introduction}
We  investigate 
the law that governs the 
power  needed to force a motion of a planar interface between two different phases of a given ferromagnetic sample with a prescribed speed $V.$
The evolution of a macroscopic phase boundary can be related rigorously to a lattice model of Ising-spins with Glauber dynamics by a multi-scale procedure, see \cite{DOPT,KS}. First, a spatial scaling of the order of the (diverging) interaction range of the Kac-potential is applied to obtain a deterministic limit on the so-called mesoscale, which follows a nonlocal evolution equation, see \cite{DOPT, CE}. This equation is then rescaled diffusively to obtain the macroscopic evolution law, in this case motion by mean curvature. For an appropriate choice of the parameters both
limits can be done simultaneously to obtain a macroscopic (and deterministic) evolution law for the phase boundary, in this case motion by mean curvature. It is natural to ask for the corresponding large deviations result, i.e., for the probability of macroscopic interfaces evolving differently from the deterministic limit law. 
This is particularly interesting when studying metastable phenomena of transitions from one local equilibrium to
another as one needs to quantify such large deviations which cannot be captured by the deterministic evolution
(for the present context of Glauber dynamics and Kac potential we also refer to
\cite{KM}).
For the first step, i.e., deviations from the limit equation on the mesoscale, this has been achieved by F. Comets, \cite{comets}.
In the present and the companion \cite{b} paper
we extend this result and derive the probability of large deviations for the macroscopic limit evolution starting from the microscopic Ising-Kac model. The technical difficulties are related to the fact that almost all of the system will be in one of the two phases, i.e., contribute zero to the large deviations cost, while
a deviation happens only at the interface. This means that the 
exponential decay rate of the probability of our events is smaller than the number of random variables involved.
As a consequence of these difficulties, our final result holds in one dimension only (i.e. no curvature), while several partial results do not depend on the dimension.
If we were to follow 
the technique used in \cite{comets} we would obtain errors which
are either diverging in a further parabolic rescaling or they can not be explicitly quantified
with respect to the small parameter.
Therefore, in this paper we use a different technique by introducing coarse-grained time-space-magnetization 
boxes and explicitly quantifying all possible transitions in the coarse-grained state space.

Let us explain more precisely the setting of this paper. We fix a space-time ($\xi,\tau$) scale (macroscopic) and we consider 
the particular 
example of an interface which is forced to move from a starting
position $\xi=0$ (at $\tau=0$) to a final position $\xi=R$
within a fixed time $T$.
If such a motion occurs with constant velocity, being $V=R/T$,
linear response theory and Onsager's principle
suggest that the power (per unit area) needed is given by $V^2/\mu$,
where $\mu$ is a mobility coefficient.
Our goal is to verify the limits of validity of this law in a stochastic model
of interacting spins which mesoscopically gives rise to a model of interfaces.

In \cite{ddp} the same question has been studied starting with a model in the
mesoscopic scale $(x,t)$ and examining the motion of the interface in the macroscopic
scale after a diffusive rescaling: $x=\epsilon^{-1}\xi$ and $t=\epsilon^{-2}\tau$, where
$\epsilon$ is a small parameter eventually going to zero.
The authors considered a non local evolution equation obtained as a gradient flow of a
certain functional penalizing interfaces.
An interface can be described as a non-homogeneous stationary solution of this equation, therefore in order to produce orbits where the interface is moving
(i.e., non stationary)
the authors included an additional
external force.
To select among all possible forces they considered as a {\it cost functional} 
an $L^2$-norm of the external force
whose minimizer provides the best mechanism for the motion of the interface.
However, in our case of starting from a microscopic model of spins, instead of postulating
an action functional we actually derive it as a large deviations functional.
Then, in order to find the best mechanism for the macroscopic motion of the interface
one has to study its minimizers. 
This is addressed in the companion paper
\cite{b} where we use a strategy
closely related to the one in \cite{ddp} but with the extra complication that the new
functional turns out to give a softer penalization on deviating profiles than
the $L^2$ norm considered in \cite{ddp}.

There is a significant number of works in the literature studying closely related problems,
mostly in the context of the stochastic Allen-Cahn equation.
In \cite{KORV, KRT, MR}, the authors study a minimization problem over all possible ``switching paths'' 
related to the  Allen-Cahn equation: The cost functional is the $L^2$-norm of the forcing in the Allen-Cahn 
equation, which is what one would heuristically expect if one could define the large deviations rate functional for 
the Allen-Cahn equation with space-time white noise. Their results deal with the meso-to-macro limit of those 
rate functionals, but do not connect these rigorously to a stochastic process on the microscale.
On the other hand, the large deviations have been studied in \cite{FJ, JM, HW}.
Furthermore, combining the above results, 
the large deviations asymptotics under diffusive rescaling of space and time are obtained
in \cite{BBP2} (see also the companion paper \cite{BBP1}): the authors consider
coloured noise and take both the intensity and the spatial correlation length of the noise to zero 
while doing simultaneously the meso-to-macro limit. This double limit is similar in spirit with our work
with the difference that our noise is microscopic and the ``noise to zero'' limit is replaced by a ``micro-to-meso''
limit. However, they state the large deviations principle directly in the $\Gamma$-limit while we only obtain
quantitative estimates for the upper and lower bound which are valid in this macroscopic scale; hence it
would be interesting as a future work to consider this analysis also in our case, maybe in higher dimensions
as well.

\section{The model and preliminary results.}\label{sec2}

\subsection{The microscopic model.}
Let $\Lambda=[-L,L]$ and $\mathcal T= [0,T]$ be the macroscopic space and time domain, respectively. 
For $\epsilon$ a small parameter we denote by 
$\Lambda_{\epsilon}=[-\epsilon^{-1}L,\epsilon^{-1}L]$
and $\mathcal{T}_{\epsilon}=[0,\epsilon^{-2}T]$
the corresponding mesoscopic domains.
Choosing another small parameter $\gamma$, we consider the microscopic lattice system
$\mathcal{S}_{\gamma}=\Lambda_{\epsilon}\cap\gamma\mathbb{Z}$,
 as viewed from the mesoscale.
We consider 
\begin{equation}\label{epsil}
\epsilon\equiv\epsilon(\gamma)=|\ln\gamma|^{-a}, 
\end{equation}
for some $a>0$ to be determined in Section~\ref{derivation}.
Let $\sigma$ be the
spin configuration $\sigma:=\{\sigma(x)\}_{x\in\mathcal{S}_{\gamma}}\in\{-1,+1\}^{\mathcal{S}_{\gamma}}$.
The spins interact via a Kac potential which depends on the parameter $\gamma$ and has the form
\[
J_{\gamma}(x,y)=\gamma J(x-y), \quad x,y\in\mathcal{S}_{\gamma},
\]
where  $J$ is a function such that
$J(r)=0$ for all $|r|>1$, $\int_{\mathbb R}J(r)dr=1$ and $J\in C^2(\mathbb{R})$.
Given a magnetic field $h\in\mathbb R$, we define the
energy of the spin configuration $\sigma_{\Delta}$ (restricted to a subdomain 
$\Delta\subset\mathcal{S}_{\gamma}$), 
given the configuration $\sigma_{\Delta^c}$ in its complement, by
\begin{equation}\label{energy}
H_{\gamma, h}(\sigma_{\Delta};\sigma_{\Delta^c})=-h\sum_{x\in\Delta}\sigma_{\Delta}(x)-\frac{1}{2}\sum_{x\neq y\in\Delta}
J_{\gamma}(x,y)\sigma_{\Delta}(x)\sigma_{\Delta}(y)-\frac{1}{2}\sum_{\substack{x\in\Delta \\ y\in\Delta^{c}}}J_{\gamma}(x,y)\sigma_{\Delta}(x)\sigma_{\Delta^c}(y).
\end{equation}
In $\mathcal S_\gamma$, we consider Neumann boundary conditions for the spins.
The corresponding finite volume Gibbs measure is given by
\begin{equation}\label{gibbs}
\mu^{\bar\sigma}_{\beta,\Delta,\gamma,h}(d\sigma)=\frac{1}{Z_{\beta,\Delta,h}}e^{-\beta H_{\gamma,h}(\sigma;\bar\sigma)},
\end{equation}
where $\beta$ is the inverse temperature and $Z_{\beta,\Delta,h}$ the normalization (partition function).
We introduce the Glauber dynamics, which satisfies the detailed balance condition with respect to the Gibbs measure defined above,  in terms of a continuous time Markov chain.
Let $\lambda:\{-1,+1\}^{\mathcal{S}_{\gamma}}\to\mathbb{R}_+$ be a bounded function and
$p(\cdot,\cdot)$ a transition probability on $\{-1,+1\}^{\mathcal{S}_{\gamma}}$ that vanishes on the diagonal:
$p(\sigma,\sigma)=0$ for every $\sigma\in\{-1,+1\}^{\mathcal{S}_{\gamma}}$.
Consider the space 
\begin{equation}\label{samplespace}
X=(\{-1,+1\}^{\mathcal{S}_{\gamma}},\mathbb{R}_+)^{\mathbb{N}},
\end{equation}
endowed with the Borel $\sigma$-algebra 
that makes the variables $\sigma_n\in \{-1,+1\}^{\mathcal{S}_{\gamma}}$ and $\tau_n\in\mathbb R_+$ measurable.
For each $\sigma\in\{-1,+1\}^{\mathcal{S}_{\gamma}}$, let $P_{\sigma}$ be the probability 
measure under which (i) $\{\sigma_n\}_{n\in\mathbb N}$, is a Markov chain with transition probability $p$
starting from $\sigma$
and (ii) given $\{\sigma_n\}_{n\in\mathbb{N}}$, the random variables $\tau_n$ 
are independent
and distributed according to an exponential law of parameter $\lambda(\sigma_n)$.
Any realization of the process can be described in terms of the infinite sequence of pairs $(\sigma_n,t_n)$ where $t_0=0$ and $t_{n+1}=t_{n}+\tau_{n}$ determining the state into which the process jumps and the time at which the jump occurs:
\[
\{\sigma_t\}_{t\ge 0}\leftrightarrow ((\sigma_1,t_1)\,,\,(\sigma_2,t_2)\,,\,\ldots\,,(\sigma_k,t_k)\,,\,\ldots).
\]
The space of realizations of the Glauber dynamics is also equivalent to $D(\mathbb{R}_+,\{-1,+1\}^{\mathcal{S}_{\gamma}})$, namely the Skorohod space of cadlag trajectories (continuous from the right and with limits from the left).

From \cite{KL} we have that for every $P_{\sigma}$ the sequence $(\sigma_n,t_n)$
is an inhomogeneous Markov chain with infinitesimal
transition probability given by
\begin{equation}\label{transition}
P(\sigma_{n+1}=\sigma',\,t\le t_{n+1}<t+dt\,|\,\sigma_{n}=\sigma ,\,t_{n}=s)=
p(\sigma,\sigma ')\lambda(\sigma)e^{-\lambda(\sigma)(t-s)}\mathbf 1_{\{t>s\}}dt.
\end{equation}
The flip rate $\lambda$ is given by
\[
\lambda(\sigma)=\sum_{x\in\mathcal{S}_{\gamma}}c(x,\sigma)
\]
and the transition probability by
\[
p(\sigma,\sigma')=[\lambda(\sigma)]^{-1}\sum_{x\in\mathcal{S}_{\gamma}}c(x,\sigma)\mathbf 1_{\sigma'=\sigma^{x}},
\]
where $\sigma^{x}$ is the configuration obtained from $\sigma$ flipping the spin located at $x$.
The flip rates $c(x,\sigma)$ for single spin at $x$ in the configuration $\sigma$ are defined by
\begin{equation}
c(x,\sigma)=\frac{1}{Z_{\gamma}(\sigma_{x^c})}e^{-\frac{\beta}{2}\Delta_x H_{\gamma}(\sigma)},\,\,\,\,\,\,\,
\Delta_x H_{\gamma}(\sigma)=H_{\gamma}(\sigma^x)-H_{\gamma}(\sigma)
=2\sigma(x)\sum_{y\neq x}J_{\gamma}(x,y)\sigma(y),
\end{equation}
where
\[
Z_{\gamma}(\sigma_{x^c})=e^{-\beta h_{\gamma}(x)}+e^{\beta h_{\gamma}(x)},\,\,\,\,\,\,\,
h_{\gamma}(x)=\sum_{y\neq x}J_{\gamma}(x,y)\sigma(y).
\]
For later use we also express the rates as:
\begin{equation}\label{rates}
c(x;\sigma)= F_{\sigma(x)}(h_\gamma(x)),\,\,\,\,\,\text{where}\,\,\,\,\,
F_{\sigma(x)}(g)=\frac{e^{-\sigma(x)\beta g}}{e^{-\beta g}+e^{\beta g}}.
\end{equation}
Note that the flip rate is bounded both from above and below:
\begin{equation}\label{boundsonrate}
c_m:=\frac{e^{-2\beta\|J\|_{\infty}}}{e^{2\beta\|J\|_{\infty}}+e^{-2\beta\|J\|_{\infty}}}
\leq c(x,\sigma)\leq \frac{e^{2\beta\|J\|_{\infty}}}{e^{2\beta\|J\|_{\infty}}+e^{-2\beta\|J\|_{\infty}}}=:c_M.
\end{equation}

\subsection{The mesoscopic model}

For $x\in\mathcal{S}_{\gamma}$,
we divide $\Lambda_{\epsilon}$ into intervals $I_{i}$, of equal length 
\[
|I_{i}|=|I|:=|\ln\gamma|^{-b}, \quad i\in \mathcal I :=\left\{-\left\lfloor\frac{\epsilon^{-1}L}{|I|}\right\rfloor,\ldots,\left\lfloor\frac{\epsilon^{-1}L}{|I|}\right\rfloor-1\right\}
\]
for some $b>0$ to be determined in Section~\ref{derivation}.
Denoting also by $I(x)$ the interval that contains the microscopic point $x\in\mathcal S_{\gamma}$,
we consider the block spin transformation given by
\begin{equation}\label{magn}
m_{\gamma}(\sigma;\,x,t)=\frac{1}{\gamma^{-1}|I(x)|}\sum_{y\in I(x)\cap\mathcal S_{\gamma}}\sigma_t(y).
\end{equation}
In the sequel we will also need to specify it by the index $i\in\mathcal I$ of the coarse cell, i.e., denote it by 
$m_{\gamma}(\sigma;i,t)$ or use a time independent version $m_{\gamma}(\sigma;i)$ as well.

In \cite{DOPT} it has been proved that as $\gamma\to 0$ the function $m_{\gamma}(\sigma;\,x,t)$
converges in a suitable topology to $m(x,t)$ which is the solution of the following nonlocal evolution equation
\begin{equation}\label{eqn}
\frac{d}{dt}m=-m+\tanh\{\beta(J\ast m)\},
\end{equation}
where $J\ast m(x)=\int_{\mathbb R} J(x-y)m(y)\, dy$.
Furthermore, this equation is related to the gradient flow of the free energy functional
      \begin{equation}
    \label{n2.6}
 \mathcal F(m)= \int_{\mathbb R} \phi_\beta(m) dx + \frac 14
 \int_{\mathbb R\times \mathbb R} J(x,y)[m(x)-m(y)]^2dx\,dy,
     \end{equation}
where $\phi_\beta(m)$ is the ``mean field excess free energy''
      \begin{equation*}
\phi_\beta(m) = \tilde\phi_{\beta}(m) - \min_{|s|\le 1}
\tilde\phi_{\beta}(s),\qquad \tilde \phi_{\beta}(m) =
-\frac{m^2}{2} - \frac 1 \beta {\mathcal S}(m),\qquad \beta>1,
     \end{equation*}
     and ${\mathcal S}(m)$ the entropy:
      \begin{equation*}
 {\mathcal S}(m)= - \frac{1-m}{2} \ln\,\frac{1-m}{2} -
\frac{1+m}{2}\ln\,\frac{1+m}{2}.
     \end{equation*}
We also define
\begin{equation}\label{gradient}
f(m):=\frac{\delta\mathcal{F}}{\delta m}=-J\ast m+\frac{1}{\beta}\arctanh m.
\end{equation}
Thus $\mathcal F$ is a Lyapunov functional for the equation (\ref{eqn}):
\[
\frac{d}{dt}\mathcal F=-\frac{1}{\beta}\int (- \beta J\ast m+\arctanh m)(m-\tanh (\beta J\ast m))\, dx\leq 0,
\]
since the  two factors inside the integral have the same sign.
This structure will be essential in the sequel.

Concerning the stationary solutions of the equation (\ref{eqn}) in $\mathbb R$, it has been proved that 
the two constant functions  $m^{(\pm)}(x):= \pm m_\beta$, with
$m_\beta>0$  solving the mean field equation $\dis{m_\beta=
\tanh\{\beta m_\beta\}}$ are stable stationary
solutions of \eqref{eqn} and are interpreted as the two pure
phases of the system with positive and negative magnetization.

Interfaces, which are the objects of this paper, are made up from particular 
stationary solutions of \eqref{eqn}. Such solutions, called {\it instantons}, exist for any
$\beta>1$ and we denote them by $\bar
m_\xi(x)$, where $\xi$ is a parameter called the center of the instanton. 
Denoting $\bar m:=\bar m_0$, we have that
     \begin{equation}
          \label{transl}
\bar m_\xi(x)=\bar m(x-\xi).
     \end{equation}
The instanton $\bar m$ satisfies
     \begin{equation}
     \label{instanton}
\bar m (x)= \tanh\left\{\beta J *\bar m (x)\right\}, \quad x\in
\mathbb R
     \end{equation}
It is an increasing, antisymmetric function which converges
exponentially fast to $\pm m_\beta$ as $x\to \pm \infty$, see e.g. 
\cite{DOPT5}, and 
there are $\alpha$ and $a$ positive so that
     \begin{equation}
     \label{1.222}
\lim_{x\to \infty} e^{\alpha x}\bar m'(x)=a,
     \end{equation} see \cite{DOP}, Theorem 3.1.
Moreover,  any other solution of \eqref{instanton} which is
strictly positive [respectively negative] as $x\to \infty$
[respectively $x\to - \infty$], is  a translate of $\bar
m(x)$, see \cite{DOPT6}.
Note also that in the case of finite volume $\Lambda_{\epsilon}$ 
the solution $\bar m^{(\epsilon)}$
with Neumann boundary conditions is close to $\bar m$ as $\epsilon\to 0$, see
\cite{BDP3}, Section 3.

\subsection{The macroscopic scale.}

This consists of the rescaled space-time domain $\Lambda\times\mathcal T$.
The corresponding profiles are rescaled versions of the functions in the mesoscopic domain.
In particular, the mesoscopically diffuse instanton is now a sharp interface between the two phases.

\subsection{The problem}

\subsubsection{Large deviations at the macroscopic scale.}
We consider an instanton initially at a macroscopic position $0$ and move
it to a final position $R$
within a fixed time $T=R/V$, where $V$ is a given value of the average velocity.
At the mesoscopic scale functions that satisfy the above requirement are profiles
in the set $\mathcal{U}[\epsilon^{-1}R,\epsilon^{-2}T]$
where
\begin{equation}\label{set}
\mathcal{U}[r,t]=\{
\phi\in C^{\infty}(\mathbb{R}\times(0,t);(-1,1)):\,
\lim_{s\to 0^+}\phi(\cdot,s)=\bar{m},\,
\lim_{s\to t^-}\phi(\cdot,s)=\bar{m}_r
\}.
\end{equation}
Due to the stationarity of $\bar{m}$, no element in $\mathcal{U}[\epsilon^{-1}R,\epsilon^{-2}T]$
is a solution of the equation \eqref{eqn}. 
In order to produce such a motion, 
in \cite{ddp} the authors considered an external force to the equation \eqref{eqn}.
Then, the optimal motion
of the interface can be found by
minimizing an
appropriately chosen cost functional.
Following their reasoning, given
a profile $\phi(x,t)$ in \eqref{set} with time derivative $\dot{\phi}(x,t)$, 
we define the following quantity:
\begin{equation}\label{functionb}
b(x,t):=\dot{\phi}(x,t)+\phi(x,t)-\tanh(\beta J\ast\phi(x,t))
\end{equation}
and we suppose that the profiles under investigation are solutions of equation \eqref{eqn} 
with additional external force $b$:
\begin{equation}\label{forcedeqn}
\dot{m}=-m+\tanh(\beta J\ast m)+b.
\end{equation}
In \cite{ddp} the cost functional has been chosen  to be $\int_{0}^{\epsilon^{-2}T}\|b(\cdot,t)\|_{L^{2}}^2 dt$.
In the present paper we derive such an action functional by considering 
the underlying microscopic process and studying the probability of observing such a deviating event.
Note that this is a large deviations away from a typical
profile that satisfies the mesoscopic equation \eqref{eqn}.
The problem is formulated as follows: 
show that the probability of the event under investigation
\begin{equation}\label{Agamma}
\{\sigma_t:\,\sigma_0\sim\bar{m}_0,\,
\sigma_{\epsilon^{-2}T}\sim\bar{m}_{\epsilon^{-1}R}\},
\end{equation}
is logarithmically equivalent to the minimal cost computed over the class $\mathcal{U}[\epsilon^{-1}R,\epsilon^{-2}T]$ 
as $\gamma\to 0$. Here we are using the symbol $\sim$ to denote a suitable notion of distance that will be 
formally given below in Definition~\ref{topology}.
In \cite{comets} the probability for the transition from the neighborhood of a stable
equilibrium to another has been studied by establishing the equivalent to the
Freidlin-Wentzell estimates, see \cite{FV}.
The corresponding cost functional for $\mathbb T\times[0,T]$ is given by
\begin{equation}\label{cost_comets}
I_{[0,T]\times\mathbb T}(\phi)
=\int_0^{T}\int_{\mathbb{T}} \mathcal H(\phi,\dot{\phi})(x,t)\,dx\,dt,
\end{equation}
where
\begin{eqnarray}\label{cost_density}
\mathcal H(\phi,\dot{\phi}) & := &
\frac{\dot{\phi}}{2}\left[
\ln\frac{
\dot{\phi}+\sqrt{(1-\phi^2)(1-\tanh^2(\beta J\ast\phi))+\dot{\phi}^2}
}{
(1-\phi)\sqrt{1-\tanh^2(\beta J\ast\phi)}
}
-\beta J\ast \phi
\right]\nonumber\\
&&
+\frac{1}{2}\left[
1-\phi\tanh(\beta J\ast\phi)-\sqrt{(1-\phi^2)(1-\tanh^2(\beta J\ast\phi))+\dot{\phi}^2}
\right].
\end{eqnarray}
However, in our case, we have to 
perform the same task but for the rescaled time and space domain
$\Lambda_{\epsilon}\times\mathcal T_{\epsilon}$ in order to obtain a result
which is valid also at the macroscopic scale.
This is technically challenging as,
in the case the time horizon as well as the volume scale with $\epsilon(\gamma)$, 
the error estimates providing \eqref{cost_comets}
are not bounded
when $\gamma\to 0$.
To overcome it, we follow a different approach
by coarse-graining
the space of realizations of the process
in all time, space and magnetization coordinates.
Then, in order to calculate the probability of an event we intersect it with all possible 
coarse-grained ``tubelets''. The final result comes from an explicit calculation of the
probability of such a tubelet and agrees with \eqref{cost_comets}.

\subsubsection{Properties of the cost functional}
Given $(\phi,\dot{\phi})$ we define
\begin{eqnarray*}
&& u:=\phi\\
&& w:=-\tanh(\beta J\ast\phi)\\
&& b:=\dot{\phi}+\phi-\tanh(\beta J\ast\phi)
\end{eqnarray*}
Then after a simple manipulation we can write $\mathcal{H}$ in the following form
(committing a small abuse of notation):
\begin{eqnarray*}
\mathcal{H}(b,u,w) & = & \frac{1}{2}\left\{
(b-u-w)\ln\frac{b-u-w+\sqrt{(b-u-w)^2+(1-u^2)(1-w^2)}}{(1-u)(1-w)}\right.\\
& & \mbox{} \left.
-\sqrt{(b-u-w)^2+(1-u^2)(1-w^2)}+1+u w
\right\}.
\end{eqnarray*}

It is a straightforward calculation to see that uniformly on $u\in[-1,1]$ and $w\in(-1,1)$ we have:
\[
\lim_{|b|\to\infty}\frac{\mathcal{H}(b,u,w)}{|b|\ln(|b|+1)}=
\frac{1}{2}\qquad\mathrm{and}
\qquad
\lim_{|b|\to 0}\frac{\mathcal{H}(b,u,w)}{b^2}=
\frac{1}{4(1+uw)}.
\]

Note that the cost assumed in \cite{ddp} is approximating the case that $b$ is small, hence it
gives a stronger penalization of the deviating profiles than the one derived
from the microscopic system.

For further properties we refer the reader to \cite{comets}. In particular, in the sequel we will
use the fact that 
\begin{equation}\label{properties}
I_{\Lambda_{\epsilon}\times\mathcal T_{\epsilon}}(\phi)<\infty\qquad\text{iff}\qquad
\dot{\phi}\ln|\dot{\phi}|, \dot{\phi}\ln\frac{1}{1-\phi}\mathbf 1_{\{\dot{\phi}>0\}},
\dot{\phi}\ln\frac{1}{1+\phi}\mathbf 1_{\{\dot{\phi}<0\}}\in L^{1}(\Lambda_{\epsilon}\times\mathcal T_{\epsilon}).
\end{equation}

The minimizers of $I_{\Lambda_{\epsilon}\times\mathcal T_{\epsilon}}(\phi)$
over the class $\mathcal{U}[\epsilon^{-1}R,\epsilon^{-2}T]$ is addressed
to the companion paper \cite{b}.
To get a rough idea,
the cost of a moving instanton
with $\epsilon$-small velocity, i.e 
\[
\phi_{\epsilon}(x,t)=\bar{m}_{\epsilon V t}(x),\,\,\,\,\,\,V=\frac{R}{T},
\]
is given by
\begin{equation}\label{cost_transl}
I_{\Lambda_\epsilon\times\mathcal T_\epsilon}(\phi_{\epsilon})
=\frac{1}{4}\|\bar{m}'\|^2_{L^2(d\nu)}V^2 T
\end{equation}
where $\bar{m}'$ is the derivative of $\bar{m}$ and $\|\cdot\|_{L^2(d\nu)}$
denotes the $L^2$ norm on $(\mathbb{R},d\nu(x))$ with
$d\nu(x)=\frac{dx}{1-\bar{m}^2(x)}$.
Following \cite{ddp} it can be shown that other ways to move
continuously the instanton are more expensive.
However, in such systems one can also observe the phenomenon of nucleations,
namely the appearance of droplets of a phase inside another.
In \cite{BDP} and \cite{BDP2} it has been proved that for such a profile the cost is
bounded by twice the free energy computed at the instanton so it can be comparable
to the cost \eqref{cost_transl} of the translating instanton. This will be properly stated in the
main results in the next section.

\section{Main results}\label{main}

We divide $\Lambda_\epsilon\times\mathcal{T}_\epsilon\times [-1,1]$
into space - time - magnetization boxes 
\[
I_i\times[j\Delta t,(j+1)\Delta t)
\times [-1+k\Delta,-1+(k+1)\Delta),
\]
where $i\in\mathcal I :=\left\{-\left\lfloor\frac{\epsilon^{-1}L}{|I|}\right\rfloor,\ldots,
\left\lfloor\frac{\epsilon^{-1}L}{|I|}\right\rfloor\right\}$,
$j\in\mathcal J:=\left\{0,1,\ldots,\left\lfloor\frac{\epsilon^{-2}T}{\Delta t}\right\rfloor-1\right\}$
and
$k\in\mathcal K^{\Delta}:=\{0,1,\ldots,\left\lfloor\frac{2}{\Delta}\right\rfloor-1\}$.
We choose the length to be
\begin{equation}\label{cg_size}
|I|=|\ln\gamma|^{-b},\,\, \Delta t=\gamma^c, \, c<1 \quad\text{and}
\quad 
\Delta=\Delta t\,\eta_{0},
\end{equation}
respectively, where 
\begin{equation}\label{eta}
\eta_{0}\equiv\eta_{0}(\gamma)=|\ln\gamma|^{-\lambda_{0}},
\end{equation}
for some number $\lambda_{0}>0$ to be determined later in \eqref{req4}. Note that each $I_i$ contains $\gamma^{-1}|I_i|$ many lattice sites of $S_{\gamma}$.
Given such a coarse cell, we 
define the set of all {\it discretized paths} by
\begin{equation}\label{sample}
\bar \Omega_{\gamma}:=
\left\{
a\equiv\{a_{i,j}\}_{i\in\mathcal I,j\in\mathcal J}:\,
 a_{i,j}\in \mathcal K^{\Delta}
\right\}.
\end{equation}

\begin{defin}\label{topology} 
Given $a\in\bar\Omega_\gamma$ and $\delta>0$, recalling the definition of $m_{\gamma}(\sigma;\,x,t)$ in \eqref{magn} for
some $x\in I_{i}$,
we say that $\sigma\in\{a\}_{\delta}$ if
$$
\sup_{i\in\mathcal I,\,j\in\mathcal J}
\left|m_{\gamma}(\sigma;\,x,j\Delta t)
-a_{i,j}\right|<\delta.
$$
Given a function $m\in L^\infty(\Lambda_\epsilon\times\mathcal{T}_{\epsilon} )$, we say that
$\sigma\in\{m\}_{\delta}$ if
$$
\sup_{i\in\mathcal I,\,j\in\mathcal J}
\left|m_{\gamma}(\sigma;\,x,j\Delta t)
-\frac{1}{|I_i|}\int_{I_i}m(x,j\Delta t)\,dx\right|<\delta.
$$
Similarly, for a time-independent function $m\in L^\infty(\Lambda_\epsilon)$ we denote by
$\sigma_t\in\{m\}_{\delta}$ (or $\{\sigma_t\sim m\}$ if we do not want to
specify the parameter $\delta$) the relation
$$
\sup_{i\in\mathcal I}
\left|m_{\gamma}(\sigma;\,x, t)
-\frac{1}{|I_i|}\int_{I_i}m(x)\,dx\right|<\delta.
$$

\end{defin}

Given a set $A\subset D(\mathbb{R}_+,\{-1,+1\}^{\mathcal{S}_{\gamma}})$,
to each $\sigma\in A$ we can associate an $a\in\bar\Omega_{\gamma}$ and a 
$\phi\in C^1(\Lambda_{\epsilon}\times \mathcal{T}_{\epsilon})$
such that $\sigma\in \{a\}_{\delta}$ and $\sigma\in \{\phi\}_{\delta}$, respectively. 

\begin{defin}\label{sets}
For $A\subset D(\mathbb{R}_+,\{-1,+1\}^{\mathcal{S}_{\gamma}})$, $\delta,\gamma>0$,
we define the sets
\begin{equation}\label{processtoa}
\bar\Omega_{\gamma,\delta}(A):=
\{
a\in\bar\Omega_{\gamma}:\, 
\exists \sigma\in A \; \;{\rm s.t.} \;\;\sigma\in \{a\}_{\delta}
\}
\end{equation}
and
\begin{equation}\label{processtoprofile}
\mathcal U_\delta (A) :=\{ \phi\in C^\infty(\Lambda_{\epsilon}\times \mathcal{T}_{\epsilon}):\,
\exists \sigma\in A \;\; {\rm s.t.} \;\; \sigma\in \{\phi\}_{\delta}\}.
\end{equation}
\end{defin}

The main result of this paper are the following quantitative estimates:
\begin{thm}\label{t1}
For $\gamma>0$ sufficiently small there exist $\delta_\gamma>0,$ 
$C_\gamma>0,$ $c_\gamma>0$   such that the following holds:
\begin{enumerate}
\item[(i)]
For a closed set 
$C\subset D(\mathbb{R}_+,\{-1,+1\}^{\mathcal{S}_{\gamma}})$ and for 
$\gamma>0$ small enough 
we have
\begin{equation}
\gamma\ln P(C)
\leq
-\inf_{\phi\in \mathcal U_{\delta_\gamma} (C)}
I_{\Lambda_{\epsilon(\gamma)}\times\mathcal T_{\epsilon(\gamma)}}(\phi)+C_{\gamma},
\end{equation}
with  
$\lim_{\gamma\to 0}C_\gamma=\lim_{\gamma\to 0} \delta_\gamma=0$,
where $\mathcal U_{\delta_\gamma}(C)$ is given in \eqref{processtoprofile} and the cost functional
$I_{\Lambda_{\epsilon(\gamma)}\times\mathcal T_{\epsilon(\gamma)}}(\phi)$
in \eqref{cost_comets}.
\item[(ii)]
Similarly, for an open set $O\subset D(\mathbb{R}_+,\{-1,+1\}^{\mathcal{S}_{\gamma}})$ and
for $\gamma>0$ sufficiently small, we have that
\begin{equation}
\gamma\ln P(O)
\ge 
-\inf_{\phi\in \mathcal U_{\delta_\gamma} (O)}
I_{\Lambda_{\epsilon(\gamma)}\times\mathcal T_{\epsilon(\gamma)}}(\phi)+c_{\gamma},
\end{equation}
where again $\lim_{\gamma\to 0}c_\gamma=\lim_{\gamma\to 0} \delta_\gamma=0$.
\end{enumerate}
\end{thm}

The above theorem is a quantitative version (for finite $\gamma$) of a
Large Deviation Principle (LDP) for $\gamma^{-1}\epsilon^{-1}$ many random variables
with a rate of only $\gamma^{-1}$.
Note that if we wanted to write a statement directly in the limit $\gamma\to 0$ one
should study the $\Gamma$-limit of the functional 
$I_{\Lambda_{\epsilon(\gamma)}\times\mathcal T_{\epsilon(\gamma)}}$, which might
be a delicate issue since we need to express the limiting functional over
singular functions and with the appropriate topology for
the LDP to hold.
However, we can find both the minimal value and the profiles to which it corresponds in the limit $\gamma\to 0$.
This is the context of a companion paper \cite{b} where we obtain a lower bound 
 for the cost functional $I_{\Lambda_{\epsilon(\gamma)}\times\mathcal T_{\epsilon(\gamma)}}$
on the set of profiles in $\mathcal{U}[\epsilon^{-1}R,\epsilon^{-2}T]$, see \eqref{set}.
We start with a definition.
\begin{defin}\label{macrocost}
Given $R,T>0$ and the mobility coefficient $\mu=:4\|\bar{m}'\|_{L^2(d\nu)}>0$, we define the cost corresponding to $n$ nucleations and the related translations by
\begin{equation}
    \label{n2.10}
w_n(R,T):= n  2\mathcal F(\bar m) + (2n+1) \left\{\frac{1}{\mu}
\left(\frac{V}{2n+1}\right)^2 T\right\},
     \end{equation} 
     where $V=R/T$, $\mathcal F$ is the free energy \eqref{n2.6} and $\bar m$ the instanton, given in \eqref{instanton}.
     \end{defin}
Note that the first term in \eqref{n2.10} corresponds to the cost of $n$ nucleations while the second
to the cost of displacement of $2n+1$ fronts (with the smaller velocity $V/(2n+1)$).

\begin{thm}\label{t2}
Let $P>\inf_{n\geq 0} w_n(R,T)$.
\begin{enumerate}
\item[(i)]
Then $\forall\zeta>0$ there exists an $\epsilon_1>0$ such that
$\forall\epsilon<\epsilon_1$
and for all sequences $\phi_{\epsilon}\in \mathcal{U}[\epsilon^{-1}R,\epsilon^{-2}T]$ with
\begin{equation}\label{finiteness}
I_{\Lambda_\epsilon\times\mathcal T_\epsilon}(\phi_{\epsilon})\leq P,
\end{equation}
we have:
\begin{equation}\label{tobeproved}
I_{\Lambda_\epsilon\times\mathcal T_\epsilon}(\phi_{\epsilon})\geq \inf_{n\geq 0} w_n(R,T)-\zeta,
\end{equation}
where
$w_n(R,T)$ is given in Definition~\ref{macrocost}.
\item[(ii)]
There exists a sequence 
$\phi_{\epsilon}\in \mathcal{U}[\epsilon^{-1}R,\epsilon^{-2}T]$ such that 
\[
\limsup_{\epsilon\to 0}
I_{\Lambda_\epsilon\times\mathcal T_\epsilon}(\phi_{\epsilon})\leq 
\inf_{n\geq 0} w_n(R,T).
\]
\end{enumerate}
\end{thm}
The proof of this theorem is given in the companion paper \cite{b}.
Combining the results in Theorem \ref{t1} and \ref{t2}
we obtain a corollary about the optimal macroscopic motion of the 
interface.
We start with some definitions:
from the cost \eqref{n2.10} we consider the set
\begin{equation}\label{argminw}
\tilde n(R,T):=\argmin \,w_n(R,T)
\end{equation}
which contains at most two elements. One can check that 
for certain values of $R$ and $T$, $n$ and $n+1$  nucleations have
the same cost for some $n$,
since we can get
the same minimum value 
by one nucleation less, but higher velocity of the newly created fronts.
Hence, the number of nucleations quantizes the cost. 
 Now we define the set of profiles that have
for some time $t\in\mathcal T_{\epsilon}$
at least the
optimal number of nucleations.
Given $\delta>0$ we define the following set of mesoscopic paths 
\begin{equation*}
{\mathcal M}^{\delta,\e}_{R,T}:=\left\{m\in L^\infty
(\Lambda_\epsilon\times\mathcal T_\epsilon):
\,\min_{n\in \tilde n(R,T)}\left(
\sup_{t\in\mathcal{T}_{\epsilon}}\mathcal F(m(\cdot,t))
-(2 n+1)\mathcal F(\bar m)
\right)>-\delta
\right\}
\end{equation*}
and the set of realizations
\begin{equation}\label{Adelta}
A_\gamma^{\delta}:=\left\{
\sigma:\ m_{\gamma}(\sigma;\cdot,\cdot)\in 
{\mathcal M}^{\delta,\e(\gamma)}_{R,T}
\right\}.
\end{equation}
Note also that here we assume that the nucleations occur simultaneously as this is the most efficient
way to do it, see \cite{b}.
The fact that the instanton has travelled at least $\epsilon^{-1} R$ is represented by the set
\begin{equation}\label{Cgamma}
C^{\delta}_\gamma:=\{
\sigma:\,
m_{\gamma}(\sigma;\cdot,T)< \bar m_{\epsilon^{-1}R}(\gamma^{-1}\cdot)
+\delta
\},
\end{equation}
where $\bar m_{\epsilon^{-1}R}$ is given in \eqref{transl}.
The following corollary states that if the transition happens, then it occurs 
through (at least)
the optimal number of nucleations, i.e., the path leaves the level set of the free energy.
\begin{coro}\label{thmwithAandC}
For any $\delta>0$ and for the sets $A^{\delta}_\gamma$ 
and $C^{\delta}_\gamma$
defined in \eqref{Adelta} and \eqref{Cgamma} we have:
\begin{equation}\label{statementtoprove}
\lim_{\gamma\to 0}P_{\sigma_0}
(A_\gamma^{\delta}|C_\gamma^{\delta})=1,
\end{equation}
where $P_{\sigma_0}$ denotes the law of the magnetization process starting at $\sigma_0$,
with $\sigma_0\in\{\bar m\}_{\gamma}$ as in \eqref{transl}.
\end{coro}
The proof follows from the previous results. 
The key point is that if we consider the cost corresponding to the sets
$(A_\gamma^{\delta})^c\cap C_\gamma^{\delta}$ and $C_\gamma^{\delta}$,  by using the corresponding estimates from Theorem~\ref{t1} for the closed and the open sets, we have that
\[
\inf_{\phi\in \mathcal U_{\delta_\gamma} ((A_\gamma^{\delta})^c\cap C_\gamma^{\delta})}
I_{\Lambda_{\epsilon(\gamma)}\times\mathcal T_{\epsilon(\gamma)}}(\phi)
-
\inf_{\phi\in \mathcal U_{\delta_\gamma} (C_\gamma^{\delta})}
I_{\Lambda_{\epsilon(\gamma)}\times\mathcal T_{\epsilon(\gamma)}}(\phi)
>0,
\]
since in the first set we do not include the optimal number of nucleations, hence
the cost is higher than in the second. Then, the proof follows by applying the estimates of
Theorem~\ref{t1} to the conditional probability.

\subsection{Strategy of the proof of Theorem~\ref{t1}}

Given a closed set $C\subset D(\mathbb{R}_+,\{-1,+1\}^{\mathcal{S}_{\gamma}})$ for $\Delta$
as in \eqref{cg_size}, consider the set $\bar\Omega_\gamma$. Now choose
$\delta:=\Delta/2$ and partition the sample space to get an upper bound by
restricting to $\bar\Omega_{\gamma,\delta}(C)$, given in \eqref{processtoa}.
Since we would like to work with smooth functions, we also define the
following intermediate space:
\begin{defin}\label{pcaff}
We define  by ${\rm PC}_{|I|} {\rm Aff}_{\Delta t}(\Lambda_{\epsilon}\times 
\mathcal{T}_{\epsilon})$ the
space of piecewise constant in space (in intervals of length $|I|$) and linear in time (in intervals 
of length $\Delta t$) functions. Given $a\in\bar\Omega_{\gamma}$, $\phi_a$ is the linear interpolation between the values $a(x,(j-1)\Delta t)$ and $a(x,j\Delta t)$):
\begin{equation}\label{phi0}
\phi_a(x,t):=\sum_i\mathbf{1}_{I_i}(x)\sum_j\mathbf{1}_{[(j-1)\Delta t,j\Delta t)}(t)
\left[
\frac{a_{i,j}-a_{i,j-1}}{\Delta t}t+j\cdot a_{i,j-1}-(j-1)\cdot a_{i,j}
\right].
\end{equation}
\end{defin}
With the above choices we have:
\begin{eqnarray}\label{strategy}
\gamma\ln P(C) 
& \leq & \gamma\ln \sum_{a\in \bar\Omega_{\gamma}} P(\{a\}_\delta\cap C)\nonumber\\
& \leq & \sup_{a\in \bar\Omega_{\gamma,\delta}(C)}\left\{-\sum_{i,j}\tilde f_{i,j}(a_{i,j})\right\}+\gamma\ln |\bar \Omega_\gamma| \nonumber\\
& \leq & -\inf_{a\in \bar\Omega_{\gamma,\delta}(C)}I_{\Lambda_{\epsilon(\gamma)}\times\mathcal T_{\epsilon(\gamma)}}(\phi_a,\dot{\phi}_{a})+C_\gamma\nonumber\\
& \leq & -\inf_{\phi\in\mathcal U_{\delta}(C) }I_{\Lambda_{\epsilon(\gamma)}\times\mathcal T_{\epsilon(\gamma)}}(\phi,\dot{\phi})+C_\gamma,
\end{eqnarray}
if we are able to find for a given tubelet $\{a\}_\delta$ an estimate of the form
\begin{equation}\label{probtubelet}
\gamma\ln P(\{a\}_\delta)\leq \sum_{i,j}\tilde f_{i,j}(a_{i,j})+C_\gamma.
\end{equation}
Here, $\tilde f_{i,j}(a)$ will be a discrete version of the density of the cost functional
we are after. 

In the second inequality we bounded the sum by the maximum value times its cardinality.
Denoting by $N_s$, $N_t$ and $N_m$ the number of space, time and
magnetization coarse cells, we have the following bound for the cardinality:
\begin{equation}\label{totboxes}
\left|\bar \Omega_\gamma\right|\leq N_m^{N_s\cdot N_t},\,\,\,\,\,\,\, \mbox{where} \,\,\,\,N_m\leq 2/\Delta.
\end{equation}
This gives
\begin{equation}\label{constantCgamma}
\gamma\ln|\bar \Omega_\gamma|=\gamma\frac{\epsilon^{-3}}{\Delta t |I|}\ln \frac{2}{\Delta}
\to 0,
\end{equation}
for all $c<1$,
as $\gamma\to 0$.

In order to prove \eqref{probtubelet},
in Section \ref{onee} we divide $\mathcal T_\epsilon$ into time intervals with less (respectively more) spin 
flips than a fixed number. We call these time intervals good (respectively bad).
We first show that the probability of having more than a given
number (still diverging) of bad time intervals is negligible.
In this way we partition the space of realizations by considering good and bad time intervals which we
study separately. In each case we obtain a different form of $\tilde f$.
In Section~\ref{goodtimeinterval} we study the probability of the tubelet in a good time interval and by
appropriately approximating it by a Poisson process for the number of positive and negative
spin flips we obtain a formula for the density of the cost functional
under the assumption that the fixed magnetization profiles $a$ are far enough from
their boundary values $\pm 1$. This assumption will be removed later in Appendix~\ref{moveaway} by showing that the probability of the process being close to
any profile $a$ can be approximated within some allowed error by the probability of
the process being close to another profile $\tilde a$ as above.
Another key step of the derivation of the cost in the good time intervals is to replace
the random by deterministic rates
and this is given in Section \ref{nondettodet0}.
Then, in Section~\ref{badtimeinterval} we treat the case of bad time intervals.
More specifically we first show a rough upper bound for the probability in a given
time interval which together with the estimated number of bad time intervals
shows that the bad time intervals have vanishing contribution to the cost.
We conclude with Section~\ref{derivation} 
where we prove that the discretized
sum is a convergent Riemann sum yielding the cost functional we are after.
To do that, we replace the discrete values $a$ by the corresponding profile $\phi_{a}$
and subsequently obtain the cost functional over such functions given by 
$I_{\Lambda_{\epsilon(\gamma)}\times\mathcal T_{\epsilon(\gamma)}}(\phi_{a})$
as in \eqref{cost_comets}.
Finally, in Lemma \ref{overmore} we argue that it is enough to minimize over smoother versions of such functions, i.e., we
will restrict our attention on the set
given in \eqref{processtoprofile}.
Once we have the upper bound we can look where the infimum occurs.
Then for the lower bound we pick a collection $\{a^*_{i,j}\}_{i,j}$
which corresponds to the infimum
and we bound the probability of an open set $O$ 
by the probability
of this particular profile, i.e.,
\begin{equation}\label{lb}
P(O)\geq P(\{a^*\}_{\delta}\cap O).
\end{equation}
We skip the explicit proof of the lower bound as it is a straightforward repetition of the 
steps for obtaining the upper bound, with small alterations which will be discussed throughout
the proof.

\section{Too many jumps are negligible.}\label{onee}

We distinguish two types of time intervals, namely those with 
less (we call them {\it good}) or more (we call them {\it bad}) 
spin flips than a fixed number $N$ to be a slightly larger number 
than the expected number of jumps within time $\Delta t$, i.e., we choose
\begin{equation}\label{N}
N:= \gamma^{-1}\epsilon^{-1}\Delta t\frac{1}{\eta_1},
\end{equation}
where 
\begin{equation}\label{eta1}
\eta_1\equiv\eta_1(\gamma):=|\ln\gamma|^{-\lambda_{1}},
\end{equation}
for some $\lambda_{1}>0$ to be determined in \eqref{req10}. 
For the time interval $[j\Delta t, (j+1)\Delta t)$
we denote the number of jumps within this interval by:
\[
N(\sigma_t,j)=\card\{t\in [(j-1)\Delta t,\, j\Delta t)
: \exists x \in \mathcal{S}_{\gamma}\,\,\mathrm{with}
\lim_{\tau\to t^-} \sigma_{\tau}(x)=-\sigma_t(x)
\}.
\]
We decompose the path space $X$ in \eqref{samplespace} as follows:
\[
X=\cup_{k\in\mathcal J}
\cup_{j_1<\ldots<j_k}D^{(k)}_{j_1,\ldots,j_k},
\]
where 
\[
D^{(k)}_{j_1,\ldots,j_k}=\{
N(\sigma_t,j)>N,\, j\in\{j_1,\ldots,j_k\}\,\text{and}
\,\, N(\sigma_t,j)\leq N,\,\mbox{otherwise}
\}
\]
is the set of realizations with $k$ bad time intervals, indexed
by $j_1,\ldots,j_k$.
Then for the probability in the left hand side of \eqref{probtubelet} we have:
\[
P(\{a\}_\delta)=P(\{a\}_\delta\cap \bar{D}_{\bar{k}})
+P(\{a\}_\delta\cap \bar{D}_{\bar{k}}^c),
\]
where
\[
\bar{D}_{\bar{k}}=\cup_{k>\bar{k}}\cup_{j_1<\ldots<j_k}D^{(k)}_{j_1,\ldots,j_k}.
\]
We select $\bar{k}$ such that $P(\{a\}_\delta\cap \bar{D}_{\bar{k}})$ is
negligible.
Note that
\begin{equation}\label{morethanN}
P_{\sigma_{(j-1)\Delta t}}(\{\sigma_t: N(\sigma_t,j)\geq N\})
\leq e^{-cN\ln\frac{N}{\lambda\Delta t}},
\end{equation}
where $\lambda:=\max_{\sigma}\lambda(\sigma)$. Therefore, given a configuration $\sigma_0$, we have
\begin{eqnarray}\label{morethanNsmall}
P_{\sigma_0}(\bar{D}_{\bar{k}})
&\leq &
\sum_{k>\bar{k}}\binom{\frac{\epsilon^{-2}T}{\Delta t}}{k}(
\sup_{\bar{\sigma}}P_{\bar{\sigma}}(N(\sigma_t,1)>N)
)^k\nonumber\\
& \leq &
\sum_{k>\bar{k}}\left(\frac{\epsilon^{-2}T}{\Delta t}\exp\{
-c\gamma^{-1}\epsilon^{-1}\Delta t\frac{1}{\eta_1}\ln\frac{1}{\eta_1}
\}
\right)^k\leq 
e^{\bar{k}[\ln \frac{\epsilon^{-2}T}{\Delta t}-c\gamma^{-1}\epsilon^{-1}\Delta t\frac{1}{\eta_1}\ln\frac{1}{\eta_1}]},
\end{eqnarray}
which is negligible if we
choose 
\begin{equation}\label{bark}
\bar{k}:= \frac{1}{\eta_2}\cdot\frac{1}{\epsilon^{-1}\Delta t\frac{1}{\eta_1}\ln\frac{1}{\eta_1}},
\end{equation}
for some
\begin{equation}\label{mulambda}
\eta_2\equiv\eta_2(\gamma)=|\ln\gamma|^{-\lambda_{2}},
\quad\text{with}\quad
\lambda_{1}>\lambda_{2}>0,
\end{equation} 
so that $\eta_{1}<<\eta_{2}$,
as required in Section~\ref{derivation}, formula \eqref{req5}. 
Notice that $\bar{k}\to\infty$ as $\gamma\to 0$ since $\Delta t =\gamma^{c}$ while all other parameters grow logarithmically in $\gamma$.

Thus, overall we show that the probability of having
too many bad time strips is negligible so for the upper bound
we estimate it by the
probability of the set $\{a\}_\delta\cap \bar{D}_{\bar{k}}^c$.
We have:
\begin{eqnarray}\label{mainsplit}
P(\{a\}_\delta\cap \bar{D}_{\bar{k}}^c) & = &
\sum_{k\leq\bar{k}}\sum_{j_1<\ldots<j_k}
\prod_{j\in\{j_1,\ldots,j_k\}}
P_{\sigma_{j-1}}(\sigma_{j\Delta t}\in\{a_{\cdot,j}\}_\delta,N(\sigma_t,j)>N)\times\nonumber\\
&& \prod_{j\notin\{j_1,\ldots,j_k\}}
P_{\sigma_{j-1}}(\sigma_{j\Delta t}\in\{a_{\cdot,j}\}_\delta,N(\sigma_t,j)\leq N),
\end{eqnarray}
which can be further bounded by taking the cardinality $\bar k\binom{\frac{\epsilon^{-2}T}{\Delta t}}{\bar k}$ of the sum over $k$ and $j_1<\ldots<j_k$ and then the max over $(k,\{j_1,\ldots,j_k\})$.
We call ${k^*,\{j_1^*,\ldots,j_{k^*}^*\}}$
the choice where the maximum is attained.
On the good time strips ($j\notin\{j_1^*,\ldots,j_k^*\}$) we derive a 
discrete version of the density of the cost functional.
On the other hand, on the bad time strips ($j\in\{j_1^*,\ldots,j_k^*\}$)
we obtain upper and lower bounds and show that since these are few the corresponding cost is negligible.
Note also that for the lower bound \eqref{lb} we can simply
restrict our attention on the good part $D_0^c$.

\section{Good time intervals}\label{goodtimeinterval}

In this section we compute the probability in a good time interval $[(j-1)\Delta t, j\Delta t)$.

\subsection{Coarse-grained spin flip markov process $\{\bar\sigma_t\}_{t\geq 0}$}\label{spinflip} 
We establish a new spin flip markov process $\{\bar\sigma_t\}_{t\geq 0}$ which is defined 
on the same space and in a similar fashion as $\{\sigma_{t}\}_{t\geq 0}$, but does not distinguish
among the spins of the same coarse cell $I_{i}$, $i\in\mathcal I$. 
The new
transition probability is given by
\begin{equation}\label{transition1}
\bar{P}(\sigma_{n+1}=\sigma ',\,t\le t_{n+1}<t+dt\,|\,\sigma_{n}=\sigma,\, t_{n}=s)
=\bar p(\sigma,\sigma ')\bar\lambda(\sigma)e^{-\bar\lambda(\sigma)(t-s)}\mathbf{1}_{\{t>s\}}dt,
\end{equation}
where $\bar p(\cdot,\cdot)$ and $\bar\lambda$ are given below.
Recalling the coarse-graining over space intervals $I_{i}$, $i\in\mathcal I$,
we first define the coarse-grained interaction potential
\begin{equation}\label{cgpot}
\bar J_{\gamma}(i,i'):=\frac{1}{\gamma^{-2}|I|^2}\sum_{x\in I_i,\,y\in I_{i'}}J_{\gamma}(x,y),
\,\,\,\,\,\text{where}\,\,i,i'\in\mathcal I,
\end{equation}
with $\bar J_{\gamma}(i,i)\equiv\bar J_{\gamma}(0)
:=\frac{1}{\gamma^{-1}|I|(\gamma^{-1}|I|-1)}\sum_{x,y\in I_i,x\neq y}J_{\gamma}(x,y)$.
Note also that for all $x\in I_i$ and $y\in I_{i'}$ we have the bound:
\begin{equation}\label{JbarJ}
|J_{\gamma}(x,y)-\bar J_{\gamma}(i,i')|\leq \gamma |I| \|J'\|_{\infty}
\mathbf 1_{|x-y|\leq 1}\mathbf 1_{|i-i'|\leq |I|^{-1}}.
\end{equation}
The coarse-grained rates for $x\in I_{i}$ are given by
\begin{equation}\label{newrates}
\bar c^i(x,\sigma):=\mathbf 1_{x\in I_i}(x) F_{\sigma(x)}(\bar h_{\gamma}(x)),
\end{equation}
where
\begin{equation}\label{barh}
\bar h_\gamma(x)=\mathbf 1_{x\in I_i}(x)
\sum_{i'\neq i}\bar J_{\gamma}(i,i') \sum_{y\in I_{i'}}
\sigma(y)+\bar J_{\gamma}(i,i)\sum_{y\in I_{i}}
\sigma(y).
\end{equation}
Then, the flip rate $\bar\lambda$ 
and the transition probability are respectively given by
\[
\bar\lambda(\sigma)=\sum_{i=1}^{\epsilon^{-2}T/|I|}\sum_{x\in I_{i}}\bar c^i(x,\sigma),\;\;\;\;
\bar p(\sigma,\sigma')=[\bar\lambda(\sigma)]^{-1}\sum_{i=1}^{\epsilon^{-2}T/|I|}\sum_{x\in I_i}\bar{c}^i(x,\sigma)
\mathbf{1}_{\sigma'=\sigma^{x}}.
\]
In the next lemma we compare the processes $\sigma$ and $\bar \sigma$:
\begin{lemma}\label{micro-to-c-g}
For any $a\in\bar\Omega_{\gamma}$ there exists $c>0$ such that for $\gamma>0$ small enough
\begin{equation}\label{mic-to-c-g1}
e^{-\beta 2cL\epsilon^{-1}\gamma^{-1}\Delta t\frac{1}{\eta_1} C^{*}(\gamma)}\leq
\frac{P_{\sigma_{(j-1)\Delta t}}(\sigma_{j\Delta t}
\in\{a_{\cdot,j}\}_\delta,N_j\leq N)}
{\bar P_{\sigma_{(j-1)}\Delta t}(\bar\sigma_{j\Delta t}\in\{a_{\cdot,j}\}_\delta,N_j\leq N)}
\leq  e^{\beta 2cL\epsilon^{-1}\gamma^{-1}\Delta t\frac{1}{\eta_1} C^{*}(\gamma)},
\end{equation}
where $\eta_1$ is given in \eqref{eta1} and 
$C^*(\gamma)=|I|\|J'\|_{\infty}+\gamma\|J\|_{\infty}$.
\end{lemma}
\begin{remark1}
Note that after taking $\gamma\ln()$ and considering all time intervals, 
the error in \eqref{mic-to-c-g1} is negligible as
$\frac{\epsilon^{-2}}{\Delta t}\epsilon^{-1}\frac{1}{\eta_{1}} \Delta t C^{*}(\gamma)\to 0$, as $\gamma\to 0$,
if we choose 
\begin{equation}\label{req0}
3a+\lambda_{1}-b<0.
\end{equation}

\end{remark1}
{\it Proof.}
We compare the rates of the processes $\sigma_{t}$ and $\bar\sigma_{t}$:
for any $x\in I_i$ from \eqref{JbarJ} and the properties of $F$ in \eqref{rates},
starting from the same configuration $\sigma'$
we have that there exists $c>0$ such that
\begin{eqnarray}
|c(x,\sigma')-\bar c^i(x,\sigma')|&\leq& c|h_{\gamma}(x)-\bar{h}_{\gamma}(x)|\nonumber\\
&\leq&c\bigg| \sum_{y\neq x}J_{\gamma}(x,y)\sigma'(y)-\sum_{k\neq i}\bar{J}_{\gamma}(k,i)
\sum_{y\in I_k}\sigma'(y)-\bar J_{\gamma}(0)\sum_{y\in I_{i}, y\neq x}\sigma'(y)\bigg|\nonumber\\
&\leq& c\bigg(\sum_{k\neq i}\sum_{y\in I_k}\big| J_{\gamma}(x,y)-\bar{J}_{\gamma}(k,i)\big|
+\sum_{y\in I_i,\,y\neq x} | J_{\gamma}(x,y)-\bar J_{\gamma}(0)|\bigg).\nonumber\\
\end{eqnarray}
Using \eqref{JbarJ} we obtain the error 
\begin{equation}\label{estimaterates}
|c(x,\sigma')-\bar c^i(x,\sigma')|\leq c \beta(|I|\|J'\|_{\infty}+\gamma\|J\|_{\infty})=:c\beta C^*(\gamma),
\end{equation}
which further gives that
\begin{equation}\label{estimatelambda}
|\lambda(\sigma')-\bar\lambda(\sigma')|\leq 2cL\epsilon^{-1}\gamma^{-1} \beta C^*(\gamma).
\end{equation}
Replacing it by the Radon-Nikodym
derivative between the laws of the processes $\sigma_t$ and $\bar\sigma_t$ 
(see e.g. 
\cite{KL}, Appendix 1, Proposition 2.6)
\begin{equation}\label{Radon}
\frac{dP}{d\bar P}\Bigg|_{\mathcal{F}_{t}}=\exp\Bigg\{\int_{0}^{t}[\lambda(\sigma_s)-\bar\lambda(\sigma_s)]ds-\sum_{s\leq t}\ln\frac{\lambda(\sigma_{s^-})p(\sigma_{s^-},\sigma_s)}{\bar \lambda(\sigma_{s^-})\bar p(\sigma_{s^-},\sigma_s)}\Bigg\},
\end{equation}
we obtain the upper bound $\gamma^{-1}\epsilon^{-1}C^*(\gamma)\Delta t$ for the integral in \eqref{Radon}
and $N C^*(\gamma)$, with $N$ as in \eqref{N} for the sum, which further yield
the bounds of \eqref{mic-to-c-g1}.\qed

\medskip

Let $\bar L$ be the generator of the new process $\{\bar\sigma_t\}_{t\geq 0}$.
We consider the magnetization density at each coarse cell $I_i$ of the new process $\{\bar\sigma_t\}_{t\geq 0}$\[
m_{\gamma}(\bar\sigma;i,t):=\frac{1}{\gamma^{-1}|I|}\sum_{x\in I_{i}}\bar\sigma_{t}(x),
\]
as in \eqref{magn} and (with slight abuse of notation) define
\begin{equation}\label{magnall}
m_{\gamma}(\bar\sigma)\equiv \{m_{\gamma}(\bar\sigma;i)\}_{i\in\mathcal I}.
\end{equation}
We are interested in the action of the generator on functions
$f\in L^{\infty}(X)$ which are constant on the level sets
$\{\bar\sigma\in X: \, m_{\gamma}(\bar\sigma;i)=m_i\in M,\,\forall i\in\mathcal I\}$.
Note that such functions have  
the property that $f(\bar\sigma)=g(m_{\gamma}(\bar\sigma))$, for some
$g\in L^{\infty}(M^{\mathcal I})$ and
$M:=\{-1,-1+\frac{2}{\gamma^{-1}|I|},\ldots,1-\frac{2}{\gamma^{-1}|I|},1\}$.
Then there is a Markov generator $\mathcal L$
on $L^{\infty}(M^{\mathcal I})$ such that for any $g\in L^{\infty}(M^{\mathcal I})$
and any $\bar\sigma\in X$
\begin{equation}\label{recandcg}
e^{\bar L t}f(\bar\sigma)=e^{\mathcal L t} g(m_{\gamma}(\bar\sigma)),
\end{equation}
where $f(\bar\sigma)=g(m_{\gamma}(\bar\sigma))$.
This is easy to show:
we first denote the new coarse-grained process by $m(t)\equiv\{m_{i}(t)\}_{i\in\mathcal I}$
whose generator $\mathcal L$ is given by
\begin{equation}\label{cggenerator}
\mathcal L g(m)=\gamma^{-1} |I|\sum_{i}\bigg(
\bar c_{+}(i,m)\Big[
g(m_i-\frac{2}{\gamma^{-1}|I|})-g(m_i)
\Big]+
\bar c_{-}(i,m)\Big[
g(m_i+\frac{2}{\gamma^{-1}|I|})-g(m_i)
\Big]
\Bigg),
\end{equation}
with rates:
\begin{equation}\label{cgrates}
\bar c_{\pm}(i,m):= \frac{1\pm m_i}{2}F_{\mp}(\bar h(i;m)),
\end{equation}
where, by a slight abuse of notation compared to \eqref{barh},
\begin{equation}\label{barhm}
\bar h(i;m):=\gamma^{-1}|I|\sum_{i'\neq i}\bar J_{\gamma}(i,i') m_{i'}+
\gamma^{-1}|I|\bar J_{\gamma}(0) m_{i}
\end{equation}
and
\begin{equation}\label{F}
F_{\mp}(h)=\frac{e^{\mp \beta h}}{e^{-\beta h}+e^{\beta h}}.
\end{equation}
When $f(\bar\sigma)=g(m_{\gamma}(\bar\sigma))$ then
$\bar Lf(\bar\sigma)=\mathcal L g(m_{\gamma}(\bar\sigma))$.
By induction on $n$, we have that 
$\bar L^nf(\bar\sigma)=\mathcal L^n g(m_{\gamma}(\bar\sigma))$
and expanding $e^{\bar L t}f$ in a power series, we obtain \eqref{recandcg}.

\subsection{Poisson process for the jumps.}\label{two}

To compute the probability for the coarse-grained process
we realize the coarse-grained Glauber dynamics by constructing for each 
$m_i$ two independent Poisson processes, $\underline{t}^i_{\pm}(m_i):=\{t^i_{\pm,1}(m_i)\leq \ldots\leq t^i_{\pm,n}(m_i)\leq \ldots\}$ called
``random times" and then taking the product over all $m_i\in M$ and all $i\in\mathcal I$.
Hence, we can construct the process
$m(t):=\{m_i(t)\}_{i\in\mathcal I}$, $t\geq 0$, as follows: if
at time $s\geq 0$ the process is in $m$ then it remains in $m$ until the minimum between the times $t^i_{\pm}:=\min_{n\in\mathbb{N}}\{t^i_{\pm,n}(m_i)\}$ and over all $i$ occurs. Then, for that $i$, the magnetization
$m_i$ increases (respectively decreases) by $\frac{2}{\gamma^{-1}|I|}$. The case $\min_i t^i_{-}=\min_i t^i_{+}$ has probability 0.

\subsection{From random to deterministic rates}\label{nondettodet0}
The complication in the construction of $m(t)$ resides on the fact that we need to know how the random times
are interrelated. Furthermore, the values of $m_{i}$ and $m_{j}$ (at the two coarse-grained boxes $I_i$ and 
$I_j$, respectively) 
are correlated via the interaction potential $\bar J_\gamma$. Hence, for both of the above reasons,
the analysis would become much simpler if we made the intensities of the random times 
independent of the current value $m_i$. To this end, we 
make them depend on some deterministic
value of the profile which remains close to $m_i$ during the whole time interval of length 
$\Delta t$. As a result, there will be
only two rates for each $i\in\mathcal{I}$: 
one for the plus jumps and the other for the minus jumps. 
Let $N_{i,j-1}^{\pm}$ be the number of plus/minus random times during the time interval 
$[(j-1)\Delta t,j\Delta t]$ that occur in the $i$-th space interval.
Note that for simplicity in the notation, in $N_{i,j-1}^{\pm}$ we do not carry the dependence on $\Delta t$.
Then, the change of the magnetization in any time interval $[(j-1)\Delta t,j\Delta t)$ 
is equal to $2(N_{i,j-1}^{-}-N_{i,j-1}^{+})$. 
To formulate this idea we introduce new deterministic rates depending on the fixed configuration 
$a\equiv\{a_{i,j}\}_{i,j}$:
\begin{equation}\label{deterministic}
\bar c_{\pm}(i,a):=\bar k_{\pm}^{i,j-1}(a_{i,j-1})F_{\mp}\left(\frac{1}{|I|}\int_{I_i}dr \,J*a_{j-1}(r)\right),
\end{equation}
where $F_{\mp}$ is given in \eqref{F},
\[
a_{j-1}(r):=\sum_{k\in\mathcal I}\mathbf{1}_{I_k}(r)a_{k,j-1},\,\,\,\, r\in\mathbb R, \,\, \,\ j\in\mathcal J
\]
and
\begin{equation}\label{k1,2}
\bar k_{\pm}^{i,j-1}(x):=\frac{1\pm x}{2}.
\end{equation}

Our goal is to use the distribution of the random variable $2(N_{i,j-1}^{-}-N_{i,j-1}^{+})$. 
More precisely, in Lemma \ref{uptocycles} below, we show that the law of two independent Poisson processes with deterministic intensities $\gamma^{-1}|I|\bar c_{\pm}(i,a)$ is close to the 
law of two independent Poisson processes with intensities $\gamma^{-1}|I|\bar c_{\pm}(i,m((j-1)\Delta t))$.

By approximating the mean field process considering constant intensities $\gamma^{-1}|I| \bar c_{\pm}(i,a)$
(one for the plus and one for the minus species), the resulting process is independent in each space box
indexed by $i\in\mathcal{I}$. 
The Poisson probability of the occurrence of $n$ random times at a given space box
within a time interval of length $\Delta t$ is given by
\begin{equation}\label{occurrence}
\mathbb P_{\gamma^{-1}|I| \bar c_{\pm}(i,a)}(N_{i,j-1}^{\pm}=n)=e^{-\gamma^{-1}|I|\bar c_{\pm}(i,a)\Delta t}\frac{(\gamma^{-1}|I|\bar c_{\pm}(i,a)\Delta t)^n}{n!}.
\end{equation}

Given $d_{i,j-1}=\frac{a_{i,j}-a_{i,j-1}}{\Delta t}\in\mathbb R$ we consider the following event
\begin{equation}\label{Bdelta}
B^{\delta}_{i,j-1}(a):=\left\{\left|
\frac{2}{\gamma^{-1}|I|}(N_{i,j-1}^{-}-N_{i,j-1}^{+})-d_{i,j-1}\Delta t
\right|<\delta, N_{i,j}\leq N
\right\},
\end{equation}
where the random variable $N_{i,j}$ stands for the number of jumps within the time interval $[(j-1)\Delta t,j \Delta t)$ in the space interval $I_{i}$.

\begin{lemma}\label{uptocycles}
Let $\nu^i=\mathbb P_{\gamma^{-1}|I| \bar c_+(i,a)}\times \mathbb P_{\gamma^{-1}|I| \bar c_+(i,a)}$ be the law of 
the product of two independent Poisson processes with intensities $\gamma^{-1}|I|\bar c_{+}(i,a)$ and $\gamma^{-1}|I|\bar c_{-}(i,a)$, respectively. Then, for any configuration $a\in\bar\Omega_{\gamma}$
and $\delta> 0$,
we have that
\begin{equation}\label{mic-to-c-g3}
P_{m((j-1)\Delta t)}(m(j\Delta t)\in\{a_{\cdot,j}\}_\delta,N_j\leq N)
\leq e^{2c\beta L\epsilon^{-1}\gamma^{-1}\frac{1}{\eta_{1}}\Delta t (C^{*}(\gamma)+\delta)}\prod_{i\in\mathcal{I}}  \nu^{i}_{m_i((j-1)\Delta t)}(B^{\delta}_{i,j-1}(a))
\end{equation}
and
\begin{equation}\label{mic-to-c-g4}
P_{m((j-1)\Delta t)}(m(j\Delta t)\in\{a_{\cdot,j}\}_\delta,N_j\leq N)
\geq e^{-2c\beta L\epsilon^{-1}\gamma^{-1}\frac{1}{\eta_{1}}\Delta t (C^{*}(\gamma)+\delta)}\prod_{i\in\mathcal{I}}  \nu^{i}_{m_i((j-1)\Delta t)}(B^{\delta}_{i,j-1}(a)),
\end{equation}
where $C^{*}(\gamma)$ is given in \eqref{estimaterates}, $\eta_{1}$ in \eqref{eta1}
and $B^{\delta}_{i,j-1}$ in \eqref{Bdelta}
with $d_{i,j-1}=\frac{a_{i,j}-a_{i,j-1}}{\Delta t}$. Moreover, we denote by $\nu^i_{{m_i((j-1)\Delta t)}}(\cdot)$ the conditional probability of an event which starts from $m_i((j-1)\Delta t)$ at time $(j-1)\Delta t$.
\end{lemma}
\begin{remark1}
Finally, note that the error is negligible 
for the choice $\delta\equiv\delta_{\gamma}=\frac{\Delta}{2}$ with $\Delta$ as in \eqref{cg_size}, since,
after considering all time intervals,
\[
\frac{\epsilon^{-2}}{\Delta t}\epsilon^{-1}\frac{1}{\eta_{1}} \Delta t (C^{*}(\gamma)+\delta_{\gamma})\to 0,\qquad{\text when}
\,\,\,\gamma\to 0,
\]
under the requirement \eqref{req0} and the fact that $\Delta t$ (in $\delta_{\gamma}$) is a power of $\gamma$.
\end{remark1}

{\it Proof.} We consider a process $\{\bar m(t)\}_{t\geq 0}$ whose rates are constant and equal to $\gamma^{-1}|I|\bar c_{\pm}(i,a)$ as in \eqref{deterministic}. 
By comparing the rates $\bar c_{\pm}(i,m)$ and $\bar c_{\pm}(i,a)$ given in \eqref{cgrates}
and \eqref{deterministic}, respectively, we have: 
\[
|\bar c_{\pm}(i,m)-\bar c_{\pm}(i,a)|\leq
\]
\begin{eqnarray}\label{rateforlemma}
&\leq&
c\delta+
c\left|\gamma^{-1}|I|\sum_{k\neq i}\bar J_{\gamma}(i,k) m_{k}+
\gamma^{-1}|I|\bar J_{\gamma}(0)m_{i}-\frac{1}{|I|}\int_{I_i}dr \,J*a_{j-1}(r)\right|\nonumber\\
&\leq  &
c\delta+c\gamma^{-1}|I|\frac{1}{|I|}\int_{I_{i}} dr\sum_{k\neq i}a_{k}\frac{1}{|I|}\int_{I_{k}}dr'|J_{\gamma}(r,r')-\bar J_{\gamma}(i,k)|
+ \nonumber\\
&&
c\gamma^{-1}|I|\sum_{k\neq i}\bar J_{\gamma}(i,k)(a_{k}-m_{k})+
\gamma^{-1}|I| \frac{1}{|I|^{2}}\int_{I_{i}\times I_{i}} dr\,dr' |J_{\gamma}(r,r')-\bar J_{\gamma}(0)|
\nonumber\\
&\leq &
c\delta+c(\gamma^{-1}|I|\frac{1}{|I|} \gamma |I|\|J'\|_{\infty}+\delta+\gamma\|J\|_{\infty}),
\end{eqnarray}
where we have used \eqref{JbarJ} for the slightly different case, namely when $r,r'\in\mathbb R$ rather than
just on $\mathcal S_{\gamma}$.
Recalling $C^{*}(\gamma)$ from \eqref{estimaterates}, we obtain:
\begin{equation}\label{estimatelambda1}
|\lambda(m)-\bar\lambda(\bar m)|\leq 2cL\epsilon^{-1}\gamma^{-1} \beta (C^{*}(\gamma)+\delta).
\end{equation}
By using \eqref{Radon} we get
\begin{equation}\label{mic-to-c-g}
e^{-2c\beta L\epsilon^{-1}\gamma^{-1}\frac{1}{\eta_{1}}\Delta t (C^{*}(\gamma)+\delta)}\leq
\frac{P_{m{(j-1)\Delta t}}(m(j\Delta t)
\in\{a_{\cdot,j}\}_\delta,N_j\leq N)}
{P_{m{(j-1)}\Delta t}(\bar m(j\Delta t)\in\{a_{\cdot,j}\}_\delta,N_j\leq N)}
\leq e^{2c\beta L\epsilon^{-1}\gamma^{-1}\frac{1}{\eta_{1}}\Delta t (C^{*}(\gamma)+\delta)}.
\end{equation}
Furthermore, since the processes $\bar m_i$ are independent with respect to $i\in\mathcal I$, we can write \eqref{mic-to-c-g} in the following form:
\begin{eqnarray}\label{mic-to-c-g2}
P_{m{(j-1)\Delta t}}(m(j\Delta t)\in\{a_{.,j}\}_{\delta},N_j\leq N) & \leq &
e^{2c\beta L\epsilon^{-1}\gamma^{-1}\frac{1}{\eta_{1}}\Delta t (C^{*}(\gamma)+\delta)}\times\nonumber\\
&&
\!\!\!\!\!\!\!\!\!\!
\!\!\!\!\!
\prod_{i}P_{m{(j-1)}\Delta t}(\bar m_{i}(j\Delta t)\in\{a_{i,j}\}_{\delta},N_{i,j}\leq N)
\end{eqnarray}
 and similarly for the lower bound.
Last, it is easy to see that given an initial condition $m_i((j-1)\Delta t)\in\{a_{i,j-1}\}_{\delta}$,
for every element of the set
$\{\bar m_i(j\Delta t)\in \{a_{i,j}\}_{\delta},N_{i,j}\leq N\}$ corresponds only one element of 
$B^{\delta}_{i,j-1}(a)$, hence the right hand side of \eqref{mic-to-c-g2} equals
that of \eqref{mic-to-c-g3}, which
concludes the proof of the lemma.
\qed

\begin{remark1}
Note that if, instead of the definition \eqref{cgpot} for the coarse potential, we used
a different one which is also more common in the literature, e.g. see \cite{errico} formula
(4.2.5.2), namely
\begin{equation}\label{cgpot2}
\bar J_{\gamma}(i,i'):=\frac{1}{|I|^2}\int_{I_i\times I_{i'}}J_{\gamma}(r,r')dr\, dr',
\quad i,i'\in\mathcal I,
\end{equation}
then the estimate \eqref{rateforlemma} would be simpler and equal to $c\delta$.
\end{remark1}

\bigskip

The next task is the asymptotic analysis of \eqref{occurrence}. 
In the lemma below we compute the cost functional for the
Poisson process.
\begin{lemma}\label{Main}
Given a profile $a\equiv \{a_{i,j}\}_{i,j}\in \bar\Omega_\gamma$,
let $\nu^i=\mathbb P_{\gamma^{-1}|I| \bar c_+(i,a)}\times 
\mathbb P_{\gamma^{-1}|I| \bar c_-(i,a)}$ be the law of two independent Poisson processes with intensities $\gamma^{-1}|I|\bar c_{+}(i,a)$ and $\gamma^{-1}|I|\bar c_{-}(i,a)$, respectively. Then, for $d_{i,j-1}=\frac{a_{i,j}-a_{i,j-1}}{\Delta t}$
and $B^{\delta}_{i,j-1}(a)$ as in \eqref{Bdelta}, with some $\delta>0$ small, 
e.g. $\delta=\Delta t\,\eta_{0}$, with $\eta_{0}$ as in \eqref{eta},
we have:
\begin{eqnarray}\label{nu1}
\left|\frac{1}{\gamma^{-1}|I|}\ln \,\nu^i_{{m_i((j-1)\Delta t)}}(
B^{\delta}_{i,j-1}(a))-\Delta t f(\hat x_{i,j-1}^{\pm};a)\right|\leq \left(\frac{\delta}{\Delta t}\right)^{\frac{1-\alpha}{2}}\Delta t,
\end{eqnarray}
for $\alpha>0$ small and
where 
\begin{equation}\label{ffunction}
f(\hat x_{i,j-1}^{\pm};a):=h\left(\hat x_{i,j-1}^{+}\,|\,\bar c_{+}(i,a)\right)+h\left(\hat x_{i,j-1}^{-}\,|\,\bar c_{-}(i,a)\right).
\end{equation}
Furthermore,
\begin{equation}\label{functionh}
h(z|\zeta):=z\ln\left(\frac{z}{\zeta}\right)-z+\zeta
\end{equation}
and the optimal values $\hat x_{i,j-1}^{\pm}$ satisfy 
\begin{equation}\label{xhat}
\hat x_{i,j-1}^{+}\hat x_{i,j-1}^{-}=\bar c_{+}(i,a)\bar c_{-}(i,a), \qquad 
2(\hat x_{i,j-1}^{-}-\hat x_{i,j-1}^{+})=d_{i,j-1}. 
\end{equation}
\end{lemma}

\begin{remark1}
The error in \eqref{nu1} is negligible if we choose $\eta_{0}$ such that
\begin{equation}\label{req4}
\epsilon^{-3}\eta_0^{(1-\alpha)/2}\to 0,\quad\text{or}\,\,\, 3a-\frac{\lambda_0}{2}(1-\alpha)<0.
\end{equation}
Moreover,
for later use, we also consider a $\Delta t$-dependent version of $f$ in \eqref{ffunction}, namely:
\begin{equation}\label{ffunctionDeltat}
f_{\Delta t}(\hat x_{i,j-1}^{\pm};a):=h\left(\hat x_{i,j-1}^{+}\,|\,\Delta t\,\bar c_{+}(i,a)\right)+h\left(\hat x_{i,j-1}^{-}\,|\,\Delta t\,\bar c_{-}(i,a)\right).
\end{equation}
Note that for the values $\hat x_{i,j-1}^{\pm}$ given in \eqref{xhat},
the following is true:
\[
f_{\Delta t}(\hat x_{i,j-1}^{\pm};a)=f(\hat x_{i,j-1}^{\pm};a)\cdot \Delta t.
\]
\end{remark1}

\medskip

The proof of the lemma will be given in Appendix \ref{nondettodet}. 
The next step is to show that the stochastic dynamics prefer to drive the system towards 
profiles $a\in\bar\Omega_\gamma$ which are away from
the boundary values $\pm 1$.
We introduce the threshold
\begin{equation}\label{delta'}
\delta':=\Delta t\cdot\eta_3,\quad\text{with}\,\,\, \eta_3\equiv\eta_3(\gamma):=|\ln\gamma|^{-\lambda_3},\,\,\lambda_3>0,
\end{equation}
where $\lambda_3$ will be determined in \eqref{req10}
and consider the class:
\begin{equation}\label{away}
\bar\Omega_\gamma^{\delta'}:=\{a\in \bar\Omega_\gamma:|a\pm 1|>\delta'\}.
\end{equation}
In the following lemma we prove that given a profile $a\in \bar\Omega_\gamma$, we can construct a new profile $\tilde a\in \bar\Omega_\gamma^{\delta'}$ 
that the process $m$ prefers to follow with higher or comparable probability.

\begin{lemma}\label{movedprofile}
Given any profile $a\equiv \{a_{i,j}\}_{i,j}\in \bar\Omega_\gamma$ and a threshold 
$\delta':=\Delta t\cdot \eta_3$ as defined in \eqref{delta'}
where $\eta_3$ satisfies the following constraint
\begin{equation}\label{req3}
3a-\lambda_3(1-\alpha)<0,
\end{equation}
$\forall\alpha>0$ small,
there exists a profile $\tilde{a}\in \bar\Omega_\gamma^{\delta'}$ (which 
can be constructed explicitly), 
such that $|1\pm\tilde{a}|\geq\delta'$ and the following bound holds:
\[
\frac{\nu^i_{{m_i((j-1)\Delta t)}}(
B^{\delta}_{i,j-1}(a))}{\nu^i_{{m_i((j-1)\Delta t)}}(
B^{\delta}_{i,j-1}(\tilde a))}\leq e^{\gamma^{-1}|I|\Delta t\,\eta_3^{1-\alpha}}.
\]
\end{lemma}
\begin{remark1}
Note that the error is negligible if we take $\gamma\ln()$ and consider all space-time coarse-grained boxes, i.e.,
\[
\gamma\frac{\epsilon^{-3}}{|I|\Delta t}\gamma^{-1}|I|\Delta t\,\eta_3^{1-\alpha}=
\epsilon^{-3}\eta_3^{1-\alpha}\to 0,
\]
under the constraint \eqref{req3}.
\end{remark1}

The proof is given in Appendix \ref{moveaway}. 
We summarize what we have done so far: by putting together the results of Lemmas~\ref{micro-to-c-g},
\ref{uptocycles}, \ref{Main} and \ref{movedprofile}
and considering the number of all time-space coarse cells,
we have the following lower and upper bounds, for $\gamma>0$ small enough
and for some $c>0$:

 \medskip
 
 ({\bf Lower Bound}) For a profile $a\equiv \{a_{i,j}\}_{i,j}\in \bar\Omega_\gamma^{\delta '}$ we have 
 \begin{eqnarray}\label{lowersum}
 P_{\sigma_{(j-1)\Delta t}}(\sigma_{j\Delta t}
\in\{a_{\cdot,j}\}_\delta,N_j\leq N)
&\geq& e^{-2c\beta L\epsilon^{-1}\gamma^{-1}\frac{1}{\eta_{1}}\Delta t (C^{*}(\gamma)+\delta)}
e^{-c\epsilon^{-1}\gamma^{-1}\Delta t(\eta_0^{(1-\alpha)/2}+\eta_3^{(1-\alpha)})}\nonumber\\
&&
\prod_{i\in\mathcal{I}} e^{-\gamma^{-1}|I|\Delta t f(\hat x_{i,j-1}^{\pm};a)}.
\end{eqnarray}
  
 ({\bf Upper Bound}) For a profile $a\equiv \{a_{i,j}\}_{i,j}\in \bar\Omega_\gamma$, there exists a profile $\tilde a\equiv \{\tilde a_{i,j}\}_{i,j}\in \bar\Omega_\gamma^{\delta'}$ such that 
 \begin{eqnarray}\label{uppersum}
P_{\sigma_{(j-1)\Delta t}}(\sigma_{j\Delta t}
\in\{a_{\cdot,j}\}_\delta,N_j\leq N) & \leq &
e^{2c\beta L\epsilon^{-1}\gamma^{-1}\frac{1}{\eta_{1}}\Delta t (C^{*}(\gamma)+\delta)}
e^{c\epsilon^{-1}\gamma^{-1}\Delta t(\eta_0^{(1-\alpha)/2}+\eta_3^{(1-\alpha)})}\nonumber\\
&&
\prod_{i\in\mathcal{I}} e^{-\gamma^{-1}|I|\Delta t f(\hat x_{i,j-1}^{\pm};\tilde a)} .
\end{eqnarray}
Note that the error is negligible under the requirements in the corresponding lemmas.

\section{Bad time intervals}\label{badtimeinterval}

Going back to \eqref{mainsplit} and the discussion below, 
for the terms in $\{a\}_\delta\cap D_{\bar{k}}^c$
with $j\notin\{j_1^*,\ldots,j_k^*\}$ we use the formula derived in the
previous section.
On the other hand, for the terms with $j\in\{j_1^*,\ldots,j_k^*\}$ 
we consider upper and lower bounds by replacing the rates by the corresponding constant ones $c_{m}$ and $c_{M}$
as in \eqref{boundsonrate}.
Hence, for the case of the upper bound (and similarly for the lower bound), we construct 
a new process $\tilde\sigma$ which is a Markov Process with 
infinitesimal transition probability $\tilde P$ given by:
\begin{equation}\label{new}
 \tilde P(\tilde\sigma_{n+1}=\tilde\sigma ',\,t\leq t_{n+1}<t+dt\,|\,\tilde\sigma_{n}=\tilde\sigma ,\,t_{n}=s) =c_M\sum_{x\in\mathcal{S}_{\gamma}}\mathbf 1_{\tilde\sigma'=\tilde\sigma^{x}}
 e^{-c_M|\mathcal{S}_{\gamma}|(t-s)}\mathbf{1}_{\{t>s\}}dt.
 \end{equation}
In the new process we have replaced the rates by constant ones in such a way to get an upper bound.
It is easy to check that
\begin{equation}\label{uap}
P_{\sigma_{(j-1)\Delta t}}(\sigma_{j\Delta t}
\in\{a_{\cdot,j}\}_\delta,N_j> N)\leq e^{-(c_m-c_M){2\gamma^{-1}\epsilon^{-1}L\Delta t}}
\tilde P_{\sigma_{(j-1)\Delta t}}(\tilde\sigma_{j\Delta t}
\in\{a_{\cdot,j}\}_\delta,N_j> N)
\end{equation}
and
\begin{equation}\label{uap1}
P_{\sigma_{(j-1)\Delta t}}(\sigma_{j\Delta t}
\in\{a_{\cdot,j}\}_\delta,N_j> N)\geq e^{-(c_M-c_m){2\gamma^{-1}\epsilon^{-1}L\Delta t}}
\tilde P_{\sigma_{(j-1)\Delta t}}(\tilde\sigma_{j\Delta t}
\in\{a_{\cdot,j}\}_\delta,N_j> N),
\end{equation}
where $\tilde P$ is the probability of the new process $\{\tilde\sigma_{t}\}_{t\geq 0}$. 
To compute the upper and lower bounds for the new process we proceed as before and consider
the corresponding mean field process $\{\tilde m_{i}(t)\}_{i\in\mathcal I,\, t\geq 0}$ with rates given by 
\[
c_{+}(i,\tilde m)=\bar k_{+}^{i,j-1}(a_{i,j-1})c_M
\qquad\text{and}\quad
c_{-}(i,\tilde m)=\bar k_{-}^{i,j-1}(a_{i,j-1})c_m.
\]
By defining the Poisson representation of the process in a similar fashion as in subsection~\ref{two} we obtain
similar upper ($g_{1}$) and lower ($g_{2}$) bounds as in \eqref{lowersum} and \eqref{uppersum}, respectively,
where instead of $f$ we have
\begin{align}
g_{1}(\hat z^{\pm}_{i,j-1};a)&= h\left(\hat z_{i,j-1}^{-}\,|\,\gamma^{-1}|I| k_-^{i,j-1}(a_{i,j-1}) c_M\right)+h(\hat z_{i,j-1}^{+}\,|\,\gamma^{-1}|I|  k_+^{i,j-1}(a_{i,j-1})c_M)\label{g1}\\
g_{2}(\hat z^{\pm}_{i,j-1};a)&= h\left(\hat z_{i,j-1}^{-}\,|\,\gamma^{-1}|I| k_-^{i,j-1}(a_{i,j-1})c_m)+h(\hat z_{i,j-1}^{+}\,|\,\gamma^{-1}|I| k_+^{i,j-1}(a_{i,j-1})c_m\right).\label{g2}
\end{align}
Here
$\hat z^{\pm}$ are computed following the Appendix \ref{nondettodet}.
Note that we also have a rough lower bound: $g_{1,2}(\hat z^{\pm};a)\geq -c_b$ where $c_b$ is a positive constant number since $h\geq 0 $.

Now we have all the ingredients to derive the discrete version
of the cost functional in the space
$\Lambda_{\epsilon}\times \mathcal{T}_{\epsilon}$.

\section{Derivation of the cost functional}\label{derivation} 

We recall from Definition~\ref{pcaff} 
the space $\text{PC}_{|I|}\text{Aff}_{\Delta t}(\Lambda_{\epsilon}\times 
\mathcal{T}_{\epsilon})$ of all functions
\begin{equation}\label{phi}
\phi_a(x,t):=\sum_i\mathbf{1}_{I_i}(x)\sum_j\mathbf{1}_{[(j-1)\Delta t,j\Delta t)}(t)
\left[
\frac{a_{i,j}-a_{i,j-1}}{\Delta t}t+j\cdot a_{i,j-1}-(j-1)\cdot a_{i,j}
\right],
\end{equation}
which are linear interpolation between the values $a(x,(j-1)\Delta t)$ and $a(x,j\Delta t)$)
and piece-wise constant in space. 
 We also consider another function which agrees with its derivative in each open interval:
\begin{equation}\label{phidot}
\psi_a(x,t)=\sum_{i,j} \frac{a_{i,j}-a_{i,j-1}}{\Delta t} 
\mathbf{1}_{I_i}(x)\mathbf{1}_{[(j-1)\Delta t,j\Delta t)}(t).
\end{equation}
We also recall that $\{k^*,j^*_1,\ldots,j^*_{k^*}\}=\argmax_{\{k,j_1,\ldots,j_k\}}P(\{a\}_\delta\cap D_{j_1,\ldots,j_k}^{(k)})$ and for simplicity we call $J^*:=\{j^*_1,\ldots,j^*_{k^*}\}$. Then,
for $a\in \bar\Omega_\gamma$,
from \eqref{mainsplit}, \eqref{uppersum}, \eqref{uap} and \eqref{g2} 
we get 
\begin{eqnarray}\label{probofB}
P(\{a\}_\delta\cap D_{\bar k}^c)
&\leq & 
\bar{k}\binom{\frac{\epsilon^{-2}T}{\Delta t}}{k}\max_{k,j_1,\ldots,j_k}
P(\{a\}_\delta\cap D_{j_1,\ldots,j_k}^{(k)})\nonumber\\
& \leq &
e^{\frac{\epsilon^{-2}}{\Delta t}2c\beta L\epsilon^{-1}\gamma^{-1}\frac{1}{\eta_{1}}\Delta t (C^{*}(\gamma)+\delta)}
\times
e^{c\epsilon^{-3}(\eta_0^{(1-\alpha)/2}+\eta_3^{(1-\alpha)})}
\times e^{-\bar k(c_m-c_M){2\gamma^{-1}\epsilon^{-1}L\Delta t}}\times
\nonumber\\
&&\times\prod _{j\notin J^*}\prod_i  e^{-\gamma^{-1}|I|\Delta t \, f(\hat x_{i,j}^{\pm};\tilde a)+o_{\gamma}(1)} \times\prod _{j\in J^*}\prod_i  e^{-\gamma^{-1}|I|\Delta t \, g_{1}(\hat z_{i,j}^{\pm};\tilde a)+o_{\gamma}(1)},
\end{eqnarray}
for some $\tilde a \in \bar\Omega_\gamma^{\delta'}$ and $\bar k$ as in \eqref{bark}.
A similar lower bound is obtained following the same reasoning.
To have a negligible error in \eqref{probofB} we need the constraints \eqref{req0},
\eqref{req4}, \eqref{req3} and
\begin{equation}\label{req5}
\gamma\bar k\gamma^{-1}\epsilon^{-1}\Delta t \to 0
\quad\text{or}\quad
\eta_1<<\eta_2, \,\,\,\text{i.e.},\,\,\,\lambda_1>\lambda_2,
\end{equation}
which is true from the choice made in \eqref{mulambda}.

The next step is to replace $f$ by the density $H$ of the cost functional:

\begin{lemma}\label{abyphi}
For every $a\in \bar\Omega_\gamma^{\delta'}$, 
with $\delta'$ as in \eqref{delta'},
$\phi_a$ and $\psi_a$ as in \eqref{phi} and \eqref{phidot},
there is a constant $C_{\gamma}\to 0$ as $\gamma\to 0$
such that
\begin{equation}\label{est}
\|
F(\hat x^{\pm};a)-
H(\phi_a,\psi_a)
\|_{L^1(\Lambda_{\epsilon}\times \mathcal{T}_{\epsilon})}
\leq C_{\gamma}.
\end{equation}
Both $F(\hat x^{\pm};a)$ and 
$H(\phi_a,\psi_a)$ are
functions in $\Lambda_{\epsilon}\times \mathcal{T}_{\epsilon}$ 
given by:
\begin{equation}\label{Fofx}
F(\hat x^{\pm};a)(x,t):=\sum_{i\in\mathcal I}\mathbf{1}_{I_i}(x)\sum_{j\in\mathcal J}\mathbf{1}_{[(j-1)\Delta t,j\Delta t)}(t) f(\hat x_{i,j}^{\pm};a),
\end{equation}
with $f(\hat x_{i,j}^{\pm};a)$ as given in \eqref{ffunction}
and
\begin{eqnarray}\label{comets}
H(\phi_a,\psi_a) & := &
\frac{\psi_a}{2}\left[
\ln\frac{
\psi_a+\sqrt{(1-\phi_a^2)(1-\tanh^2(\beta J\ast\phi_a))+\psi_a^2}
}{
(1-\phi_a)\sqrt{1-\tanh^2(\beta J\ast\phi_a)}
}
-\beta J\ast \phi_a
\right]\nonumber\\
&&
+\frac{1}{2}\left[
1-\phi_a\tanh(\beta J\ast\phi_a)-\sqrt{(1-\phi_a^2)(1-\tanh^2(\beta J\ast\phi_a))+\psi_a^2}
\right],
\end{eqnarray}
where the $x,t$ dependence is hidden in $\phi_a$ and $\psi_a$.
\end{lemma}

{\bf Proof:}
We first estimate the difference between $\hat x^{+}_{i,j}$ as in \eqref{admissible}
and $y\equiv y(\phi_a,\psi_a)$ with
\begin{equation}\label{y}
y(\phi_a,\psi_a)=-\frac{\psi_a}{4}+\sqrt{
\frac{\psi_a ^2}{16}+c_+(\phi_a)c_-(\phi_a)}.
\end{equation}
The rates $c_{\pm}(\phi_a)$ are defined analogously to $\bar c_{\pm}(i,a)$ in \eqref{deterministic} where instead of $a_{j-1}(x)$ we have $\phi_a$, that is, 
\[
c_{\pm}(\phi_a):=\frac{1\pm\phi_a}{2}F_{\mp}\left(J*\phi_{a}\right) .
\]
By comparing to the rates $\bar c_{\pm}(i,a)$ we obtain that for 
$x\in I_i$ and  $t\in [(j-1)\Delta t, j\Delta t)$:
\begin{equation}\label{xhatminusy}
|\hat x^{+}_{i,j}-y(\phi_a,\psi_a)(x,t)|\leq c|\psi_{a}(x,t)\Delta t|^{1/2},
\end{equation}
for some $c>0$.
Moreover the following identities are satisfied by the above rates,
\[
F_{-}\left(J*\phi_{a}\right) \cdot F_{+}\left(J*\phi_{a}\right) =\frac{1}{(e^{\beta J\ast\phi_a}+e^{-\beta J\ast\phi_a})^2}=\frac{1}{4}(1-\tanh^2(\beta J\ast\phi_a))
\]
and
\[
c_+(\phi_a)+c_-(\phi_a)=\frac{1}{2}[1-\phi_a\tanh(\beta J\ast\phi_a)].
\]
From these and after some straightforward cancellations, we rewrite the function $H(\phi_a,\psi_a)$ 
in \eqref{comets} as follows:
\[
H(\phi_a,\psi_a)=h\left(y(\phi_a,\psi_a)\,|\,c_+(\phi_a)\right)+h(y(\phi_a,\psi_a)+\frac{\psi_a}{2}\,|\,c_-(\phi_a)),
\]
where $h$ is defined in \eqref{functionh}.
Notice the similarity with $f(\hat x_{i,j}^{\pm};a)$, where  $\phi_a$ and $c_{\pm}(\phi_a)$ 
have replaced $a$ and $\bar c_{\pm}(i,a)$, respectively.

 Then, for the difference $|f(\hat x_{i,j}^{\pm};a)-H(\phi_a,\psi_a)|$, it suffices to estimate the following as the other terms can be treated in a similar fashion:
\[
\left|\hat x^{+}_{i,j}\ln\frac{\hat x^{+}_{i,j}}{1+a}-y\ln\frac{ y}{1+\phi_a}\right|\leq
|\hat x^{+}_{i,j}-y|^{1-\alpha}+|\hat x^{+}_{i,j}|\cdot |\ln\frac{1+a}{1+\phi_a}|+|\hat x^{+}_{i,j}-y|\cdot |\ln (1+\phi_a)|.
\]
The first term is given in \eqref{xhatminusy} so we require that
\begin{equation}\label{req7}
\epsilon^{-3}|\dot{\phi}_{a}\Delta t|^{(1-\alpha)/2}\to 0,\qquad \text{as}\quad \gamma\to 0.
\end{equation}
Note that if all allowed spin-flips occur on the same space coarse-grained box we have the bound 
\begin{equation}\label{derivative}
|\dot{\phi}_a|\leq\frac{N}{\gamma^{-1}|I|}\leq\frac{\epsilon^{-1}}{\eta_1\cdot |I|},
\end{equation}
where $N$ were chosen in \eqref{N}.
Thus, requirement \eqref{req7} is easily satisfied since $\Delta t=\gamma^c$.

The main difficulty is in the second term since, in some regimes, $|\hat x^{+}_{i,j}|$ may be large and at the same time
$1+\phi_{a}$ small.
This occurs when the given profile $a$ (and subsequently also $\phi_{a}$) is very close to the
boundary value $-1$ (recall the lower bound $1+\phi_a\geq \Delta t\cdot\eta_3$ from \eqref{away})
with a negative derivative which can also be large in absolute value, given by \eqref{derivative}.
Due to the symmetry of the problem the same holds for the case of a profile going up and
being close to the upper boundary $+1$ in which case the ``bad" term is
$|\hat x^{-}_{i,j}|\cdot |\ln\frac{1-a}{1-\phi_a}|$.
More precisely, in \eqref{xopt}, if $\frac{d_{i,j}}{4}<-\sqrt{\frac{B(a,\Delta t)}{(\Delta t)^2}}<0$, then 
$|\hat x^{+}_{i,j}|\lesssim |\frac{d_{i,j}}{4}|\lesssim\frac{\epsilon^{-1}}{\eta_1\cdot |I|}$.
We fix a threshold 
\begin{equation}\label{eta4}
\eta_4\equiv\eta_4(\gamma):=|\ln\gamma|^{-\lambda_4},\,\,\,\lambda_4>0,
\end{equation}
such that $\eta_4>>\Delta t$
and we split the integral $\int |\hat x^{+}_{i,j}|\cdot |\ln\frac{1+a}{1+\phi_a}|\,dx\,dt$
into the set $\{1+\phi_a>\frac{\Delta t}{\eta_4}\}$ and its complement.
For the first we have that
\[
\frac{1+a}{1+\phi_a}=1+\frac{a-\phi_a}{1+\phi_a},\,\,\,\mathrm{where}\,\,\, \left|\frac{a-\phi_a}{1+\phi_a}\right|
\leq \frac{|\psi_a|\cdot\Delta t}{\frac{\Delta t}{\eta_4}}\leq\frac{\eta_4\cdot\epsilon^{-1}}{\eta_1\cdot |I|}
\]
and we choose 
\begin{equation}\label{req1}
\Delta t << \eta_4<<\epsilon\cdot\eta_1\cdot |I|.
\end{equation}
Under this condition we obtain that
\begin{equation}\label{fp}
\int_{\{1+\phi_a>\frac{\Delta t}{\eta_4}\}} |\hat x^{+}_{i,j}|\cdot \left|\ln\frac{1+a}{1+\phi_a}\right|\,dx\,dt
\leq \epsilon^{-4}\frac{\eta_4\cdot\epsilon^{-1}}{\eta_1^2\cdot |I|^2}.
\end{equation}
This is vanishing provided that
\begin{equation}\label{c1}
\eta_4<<\eta_1^2\cdot |I|^2\cdot\epsilon^4,\,\,\,\text{i.e.},\,\,\, \lambda_4>2\lambda_1+2b+4a,
\end{equation}
which also covers the previous requirement \eqref{req1}.

In the complement, recalling the properties \eqref{properties} 
of the functional, we exploit the fact that 
$\psi_a\ln(1+\phi_a) \in L^1(\Lambda_{\epsilon}\times \mathcal{T}_{\epsilon})$ for
$\psi_a=\dot{\phi}_{a}$.
Indeed, we have that:
\begin{equation}\label{measure}
P>\int_{\{1+\phi_a\leq\frac{\Delta t}{\eta_4}\}}|\psi_a|\cdot |\ln(1+\phi_a)|\,dx\,dt
>\ln\Delta t\int_{\{1+\phi_a\leq\frac{\Delta t}{\eta_4}\}}|\psi_a|.
\end{equation}
On the other hand, we also have that $\frac{1+a}{1+\phi_a}>1$ which implies that
\begin{equation}\label{small}
\left| \ln\frac{1+a}{1+\phi_a}
\right|\leq\left| \frac{1+a}{1+\phi_a}-1\right|=\left| \frac{a-\phi_a}{1+\phi_a}\right|\leq\frac{\epsilon^{-1}}{\eta_1 |I|\eta_3},
\end{equation}
from \eqref{derivative} and the fact that $|1+\phi_a|\geq\Delta t \,\eta_3$.
From \eqref{measure} and \eqref{small} we obtain:
\begin{equation}\label{sp}
\int_{\{1+\phi_a\leq\frac{\Delta t}{\eta_4}\}} |\hat x^{+}_{i,j}|\cdot \left|\ln\frac{1+a}{1+\phi_a}\right|\,dx\,dt
\leq \frac{\epsilon^{-1}}{\eta_1 |I|\eta_3}\cdot\frac{P}{\ln\Delta t},
\end{equation}
which is vanishing under the requirement that
\begin{equation}\label{c2}
|\ln\Delta t|^{-1}<<\eta_1 |I|\eta_3\cdot\epsilon,\,\,\,\text{i.e.},\,\, 1>\lambda_1+b+\lambda_3+a.
\end{equation}
It is easy to check that the requirements \eqref{req0} for $\eta_1$, \eqref{req3} for $\eta_3$
and \eqref{c2} for both, can be simultaneously satisfied, e.g. by choosing $\lambda_1$
and $\lambda_3$ such that
\begin{equation}\label{req10}
1>2\lambda_1+\frac 43\lambda_3(1-\alpha).
\end{equation}
Then the other parameters can be chosen as follows: $\eta_0$ from requirement \eqref{req4},
$\eta_2$ from \eqref{req5}
and $\eta_4$ from \eqref{c1}. The parameters $a$ and $b$, 
for $\epsilon$ and $|I|$ respectively, have more freedom, but within the limits of the above
constraints.
Finally, the error  $C_{\gamma}$ in  \eqref{est} is given by the right hand sides of \eqref{fp} and \eqref{sp} which are vanishing as $\gamma\to 0$.
\qed

\medskip
 
Putting together good and bad time intervals from \eqref{nu1} and \eqref{g1}-\eqref{g2}, we obtain  the following bound for the last two factors of \eqref{probofB}:
\begin{equation}\label{sumfg}
\exp\left\{-\gamma^{-1}\left(\sum_{i\in\mathcal I}\Big(\sum_{j\in J^*}g_{1,2}(\hat z_{i,j}^{\pm};a)+\sum _{j\notin J^*}f(\hat x_{i,j}^{\pm};a)\Big)|I|\Delta t\right)\right\}.
\end{equation}

Since both $f$ and $g_{1,2}$ are integrable functions in $\Lambda_{\epsilon}\times \mathcal{T}_{\epsilon}$
and $|J^{*}|/(\epsilon^{-2}/\Delta t)$ is negligible.
Using again Lemma \ref{abyphi} we have that
\eqref{sumfg} equals $\int_{\Lambda_{\epsilon}\times\mathcal{T}_{\epsilon}} H(\phi_a,\psi_a)dx\,dt$
plus vanishing error as $\gamma\to 0$.
We conclude the last step of the strategy \eqref{strategy}
by restricting to the class of smoother functions:
\begin{lemma}\label{overmore}
Given a closed set $C\subset D(\mathbb{R}_+,\{-1,+1\}^{\mathcal{S}_{\gamma}})$, 
for some $\delta,\delta'>0$
we denote by $\bar\Omega_{\gamma,\delta}^{\delta'}(C)$
the set of profiles in $\bar\Omega_{\gamma,\delta}(C)$ defined in \eqref{processtoa},
with the extra property that $|a\pm 1|>\delta'$.
Then, for such a profile $a\in\bar\Omega_{\gamma,\delta}^{\delta'}(C)$ and $\delta,\delta'$
chosen as before,
we have that
\begin{equation}\label{finalinf}
\inf_{a\in\bar\Omega_{\gamma,\delta}^{\delta'}(C)}
\int_{\Lambda_{\epsilon}\times\mathcal{T}_{\epsilon}} H(\phi_a,\psi_a)dx\,dt
\geq \inf_{\phi\in\mathcal{U}_\delta(C)}I_{\Lambda_{\epsilon}\times \mathcal{T}_{\epsilon}}(\phi)+C_{\gamma},
\end{equation}
 with the same $C_{\gamma}$ as in \eqref{est}.
\end{lemma}

{\it Proof.}
Mollified versions of $(\phi_{a},\psi_{a})$ are elements in $\mathcal{U}_\delta(C)$
to which we can restrict ourselves by obtaining a lower bound. 
Furthermore, mollified functions are close in $L^1$ to the original ones.
The same is true for their images under integrable functions
such as the ones in $H(\phi_a,\dot{\phi}_{a})$.
Hence, we can approximate $H(\phi_a,\dot{\phi}_{a})$ by $H$ evaluated at mollified versions
of $\phi_{a}$ with a negligible error which is similar to the one in 
Lemma~\ref{abyphi}. This is a standard calculation and
details are omitted.
\qed

\bigskip

{\bf Acknowledgments.}
We would like to thank Guido Manzi and Errico Presutti for many fruitful discussions.

\appendix

\section{Properties of the Poisson process}

In this appendix we obtain an asymptotic formula for the logarithm of the Poisson distribution.
Before proceeding with the proof of  Lemma~\ref{Main}, we establish some notation. For every $i\in\mathcal{I}$ and 
$j\in\mathcal J$, we
define the random variables
\begin{equation}\label{x}
X^{i,j-1}:=\frac{N_{i,j-1}^{+}}{\gamma^{-1}|I|}
\end{equation}
and
\begin{equation}\label{k3}
K^{i,j}:=\frac{2\big(N_{i,j-1}^{-}-N_{i,j-1}^{+}\big)}{\gamma^{-1}|I|}.
\end{equation}
Given $a\in \bar\Omega_\gamma$, we denote by $R_{i,j}^{\delta}(a)$ the range of the values that the
pair of random variables $(X^{i,j-1},K^{i,j})$ can take.
This is determined by the set $\{|K^{i,j}-d_{i,j-1}\Delta t|<\delta\}$ for $K^{i,j}$ and
by the set
$[m^{\delta}_{i,j}(K^{i,j}),M^{\delta}_{i,j}(a)]$ for $N_{i,j-1}^+$.
In the latter, we have defined
\begin{align}
m^{\delta}_{i,j}(K^{i,j})&:=\max\left\{0,-\gamma^{-1}|I|K^{i,j}\right\},\label{min}\\
M^{\delta}_{i,j}(a)&:=\gamma^{-1}|I|\cdot\left(\min\left\{\bar k^{i,j-1}_{+}(a),\bar k^{i,j-1}_{-}(a)-K^{i,j}\right\}-\delta\right),\label{max}
\end{align}
for the lower and upper limits (respectively) of the potential values of $X^{i,j-1}$, given
$d_{i,j-1}$ as in Lemma \ref{uptocycles} and $\bar k^{i,j-1}_{\pm}(a)$ in \eqref{k1,2}. 
Note that in $M^{\delta}_{i,j}(a)$ the minimum is over the number of pluses at
time $(j-1)\Delta t$ and the number of minuses at the next time $j\Delta t$, as the number of
pluses that become minuses cannot exceed neither of them.
By \eqref{occurrence} we have:  
\[
\nu^{i}_{m_i((j-1)\Delta t)}\big( B^{\delta}_{i,j-1}(a)\big)=
\]
\begin{eqnarray}\label{fullnu} 
&=& \!\!\!\!\!\!\!\!\!\sum_{(n^{-}_{i,j-1},n^{+}_{i,j-1})\,\in B^{\delta}_{i,j}} \mathbb P_{\gamma^{-1}|I| \bar c_{-}(i,a)}(N_{i,j-1}^{-}=n^{-}_{i,j-1})\,\mathbb P_{\gamma^{-1}|I| \bar c_{+}(i,a)}(N_{i,j-1}^{+}=n^{+}_{i,j-1})\nonumber\\
&=& \!\!\!\!\!\!\!\!\!\sum_{(n^{+}_{i,j-1},k^{i,j})\in R^{\delta}_{i,j}(a)}\mathbb P_{\gamma^{-1}|I| \bar c_{-}(i,a)}(N_{i,j-1}^{-}=n^{+}_{i,j-1}+\frac{\gamma^{-1}|I|k^{i,j}}{2})\,\mathbb P_{\gamma^{-1}|I| \bar c_{+}(i,a)}(N_{i,j-1}^{+}=n^{+}_{i,j-1}).
\;\;\;\;\;\;\;
\end{eqnarray}

For $n^{+}_{i,j-1}$ and $n^{+}_{i,j-1}+\gamma^{-1}|I|k^{i,j}$ large enough, we apply Stirling's formula to
\eqref{fullnu} and using \eqref{occurrence} we obtain the following expression: 
\begin{equation}\label{first-fa}
\sum_{(x_{i,j-1}^{+},k^{i,j})\in \gamma^{-1}|I|R^{\delta}_{i,j}(a)}\exp\left(-\gamma^{-1}|I| f_{\Delta t}(x_{i,j-1}^{\pm};a)+o_{\gamma}(1)\right),
\end{equation}
where $f_{\Delta t}(x_{i,j-1}^{\pm};a)$ is given in \eqref{ffunctionDeltat}
and $x_{i,j-1}^{\pm}$ represents the number of occurrence of the random times 
$N_{i,j-1}^{\pm}$ divided by $\gamma^{-1}|I|$. 
Recall also 
that $x^{-}_{i,j-1}=x^{+}_{i,j-1}+\gamma^{-1}|I|k^{i,j}$. Moreover,
note that in the latter sum, $k^{i,j}$ denotes a rescaled number by $\gamma^{-1}|I|$ 
while in the sum in \eqref{fullnu} it is not rescaled.
\medskip

 \subsection{Asymptotics of the Poisson process, proof of Lemma~\ref{Main}}\label{nondettodet}
 
We give the asymptotic analysis of the Poisson Process.
\medskip

{\it Proof of Lemma~\ref{Main}.} We optimize the exponent of \eqref{first-fa} with respect to $x_{i,j-1}^{+}\in \gamma^{-1}|I|R^{\delta}_{i,j}(a)$ (viewing $k^{i,j}$ as a parameter)
and using the fact that $x^{-}_{i,j-1}=x^{+}_{i,j-1}+\gamma^{-1}|I|k^{i,j}$.
The optimal value is given by
\begin{equation}\label{xopt}
x^{+,\rm{opt}}_{i,j-1}=
-A(k^{i,j})+\sqrt{A(k^{i,j})^2+B(a, \Delta t)}\geq 0,
\end{equation}
where
\[
A(k^{i,j})=\frac{k^{i,j}}{4}
\,\,\,\,\,
\mathrm{and}\,\,\,\,\,
B(a, \Delta t )=\bar c_{+}(i,a) \bar c_{-}(i,a)(\Delta t)^2.
\] 
Calling 
\[
\bar{A}(a,\Delta t ):=\frac{d_{i,j-1}}{4}\Delta t,
\] 
we define 
\begin{align}\label{bary}
\bar x_{i,j-1}^+&:=-\bar{A}(a,\Delta t)+\sqrt{\bar{A}(a,\Delta t )^2+B(a,\Delta t)}\nonumber\\
\;&=\Delta t \Bigg(-\frac{d_{i,j-1}}{4}+\sqrt{\frac{d_{i,j-1}^{2}}{16}+\bar c_{+}(i,a) \bar c_{-}(i,a)}\Bigg)\nonumber\\
&=:\Delta t\,\bar y_{i,j-1}(a).
\end{align}
By using the second property of the set $R^{\delta}_{i,j}(a)$, namely that $|k^{i,j}-d_{i,j-1}\Delta t|<\delta$ and comparing \eqref{xopt} and \eqref{bary} we have that:
\[
|\frac{x^{+,\rm{opt}}_{i,j-1}}{\Delta t}-\bar y_{i,j-1}(a)|
\leq \frac 12\frac{\delta}{\Delta t}+\left(\frac{\delta}{\Delta t}\right)^{1/2},
\]
which implies that 
\[
|\frac{x^{+,\rm{opt}}_{i,j-1}}{\Delta t}\ln\frac{x^{+,\rm{opt}}_{i,j-1}}{\Delta t\,\bar c_{+}(i,a)}-\bar y_{i,j-1}(a)\ln\frac{\bar y_{i,j-1}(a)}{\bar c_{+}(i,a)}| \leq
\]
\[
\leq
|\frac{x^{+,\rm{opt}}_{i,j-1}}{\Delta t}-\bar y_{i,j-1}(a)|^{1-\alpha}+|\frac{x^{+,\rm{opt}}_{i,j-1}}{\Delta t}-\bar y_{i,j-1}(a)|\cdot |\ln c_{+}(i,a)|
 \leq   \left(\frac{\delta}{\Delta t}\right)^{\frac{1-\alpha}{2}}.
\]
Thus,
\[
\left|h\left(x^{+,\rm{opt}}_{i,j-1}\,|\,\Delta t \,\bar c_{+}(i,a)\right)-\Delta t\,h\left(\bar y\,|\,c_{+}(i,a)\right)\right|\leq \left(\frac{\delta}{\Delta t}\right)^{\frac{1-\alpha}{2}}\Delta t.
\]
We treat the term $h\left(x^{+,\rm{opt}}_{i,j-1}+\frac{k^{i,j}}{2}\,|\,\bar c_{-}(i,a)\right)$ similarly. 
Thus, 
the optimal values are  
\begin{equation}\label{admissible}
\hat x_{i,j-1}^{+}:=\bar y_{i,j-1}(a)\quad {\rm{and}} {\quad} \hat x_{i,j-1}^{-}:=\frac{d_{i,j-1}}{2}+ \bar y_{i,j-1}(a).
\end{equation}
Thus, we substitute them in \eqref{first-fa} and since the 
cardinality of the sum is negligible after we take $\gamma\ln()$,
we conclude the proof of the lemma.\qed

\bigskip

\subsection{Move profiles away from $\pm 1$, proof of Lemma~\ref{movedprofile}}\label{moveaway}

We show that the stochastic dynamics drive the magnetization
profile away from the boundaries $\pm 1$.

\medskip
{\it Proof of Lemma~\ref{movedprofile}.}
Whenever the profile $a$ enters the safety region $ |1\pm a| \leq\delta'$
we move it away from it by $\delta'$. We define a new profile $\tilde a$ as follows:
\begin{equation}\label{tilde}
\tilde a_{i,j}:=(a_{i,j}-\delta')\mathbf{1}_{\{a_{i,j}>1-\delta'\}}+
a_{i,j}\mathbf{1}_{\{-1+\delta'\leq a_{i,j}\leq 1-\delta'\}}+(a_{i,j}+\delta')\mathbf{1}_{\{a_{i,j}< -1+\delta'\}},
\end{equation}
 with $\delta'$ as in \eqref{delta'} under the constraint \eqref{req3} and by choosing  
it to be a multiple of $\Delta$ we have that $\tilde a_{i,j}\in \bar \Omega_{\gamma}$.
Next, we consider the case when the fixed configuration $a$ is close to the $+1$ boundary, with the other
case being similar due to the symmetry of the problem.

It is more convenient to slightly change the notation for $f_{\Delta t}(x_{i,j-1}^{\pm};a)$ 
making explicit the dependence on $k^{i,j}$, i.e., writing $f_{\Delta t}((x_{i,j-1}^{+}, k^{i,j}); a)\equiv f_{\Delta t}(x_{i,j-1}^{\pm}; a)$.
Then, the strategy goes as follows:
we seek an injective map $\iota$ 
in such a way that the following two inequalities are true: \begin{eqnarray}\label{ratioofnu}
\frac{\nu^i_{{m_i((j-1)\Delta t)}}(
B^{\delta}_{i,j-1}(a))}{\nu^i_{{m_i((j-1)\Delta t)}}(
B^{\delta}_{i,j-1}(\tilde a))}&=&\frac{\sum_{(x_{i,j-1}^{+},k^{i,j})\in R_{i,j}^{\delta}(a)}\frac{e^{-\gamma^{-1}|I|f_{\Delta t}((x_{i,j-1}^{+}, k^{i,j}); a)}}{e^{-\gamma^{-1}|I|f_{\Delta t}(\iota(x_{i,j-1}^{+}, k^{i,j}); \tilde a)}}e^{-\gamma^{-1}|I|f_{\Delta t}(\iota(x_{i,j-1}^{+}, k^{i,j});\tilde a)}}{\sum_{(\tilde x_{i,j-1}^{+},\tilde k^{i,j})\in R_{i,j}^{\delta}(\tilde a)} e^{-f_{\Delta t}(\tilde x_{i,j-1}^{+}, \tilde k^{i,j}; \tilde a)}}\nonumber\\
&\leq&e^{M(\gamma)}\frac{\sum_{(x_{i,j-1}^{+},k^{i,j})\in R_{i,j}^{\delta}(a)}e^{-\gamma^{-1}|I|f_{\Delta t}(\iota(x_{i,j-1}^{+}, k^{i,j}); a)}}{\sum_{(\tilde x_{i,j-1}^{+},\tilde k^{i,j})\in R_{i,j}^{\delta}(\tilde a)} e^{-\gamma^{-1}|I|f_{\Delta t}((\tilde x_{i,j-1}^{+}, \tilde k^{i,j}); \tilde a)}}\leq e^{M(\gamma)},
\end{eqnarray}for some $M(\gamma)$ to be estimated.

\medskip

\underline{Definition of the injective map $\iota$.} 
We have three cases: suppose that the profile $a$ is close to the $+1$ boundary at time $(j-1)\Delta t$, $j\Delta t
$ or both. 
For every $(x_{i,j-1}^{+},k^{i,j})\in R_{i,j}^{\delta}(a)$
we choose a pair 
$(\tilde x_{i,j-1}^{+}, \tilde k^{i,j}):=\iota(x_{i,j-1}^{+}, k^{i,j})\in R_{i,j}^{\delta}(\tilde a)$ by replacing $d_{i,j-1}$ by
\begin{equation*}\label{primes}
\tilde d_{i,j-1}=\frac{\tilde a_{i,j}-\tilde a_{i,j-1}}{\Delta t},
\end{equation*}
with $\tilde a_{i,j-1}=a_{i,j-1}-\delta'$ or $\tilde a_{i,j}=a_{i,j}-\delta'$, respectively. 
Then,
for the first inequality of \eqref{ratioofnu},
the difference
\[
-f_{\Delta t}(x_{i,j-1}^{+}, k^{i,j}; a)+f_{\Delta t}(\iota(x_{i,j-1}^{+}, k^{i,j}); \tilde a)=
\]
\[
=-x^+\ln\frac{x^+}{c_{+}(i,a)\Delta t}+\tilde x^+\ln\frac{\tilde x^+}{ c_{+}(i,\tilde a)\Delta t}+2(x^+-\tilde x^+)+\Delta t\left(c_{+}(i,\tilde a)-c_{+}(i,a)\right)-
\]
\begin{equation}\label{difference}
-(x^++ \frac{k}{2})\ln\frac{x^++ \frac{k}{2}}{ c_{-}(i,a)\Delta t}+(\tilde x^++\frac{k}{2})\ln\frac{\tilde x^++\frac{\tilde k}{2}}{ c_{-}(i,\tilde a)\Delta t} +\frac{1}{2}(k-\tilde k)+\Delta t\left(c_{-}(i,\tilde a)-c_{-}(i,a)\right),
\end{equation}
can be estimated using the following inequalities:
\begin{equation}\label{plusinequality}
-x^+\ln\frac{x^+}{c_{+}(i, a)\Delta t}+\tilde x^+\ln\frac{\tilde x^+}{c_{+}(i,\tilde a)\Delta t}\leq
\Delta t\left|\frac{x^+-\tilde{x}^+}{\Delta t}\right|^{1-\alpha}+\tilde x^+\ln\frac{c_{+}(i, a)}{c_{+}(i, \tilde a)}+(x^+-\tilde x^+)\ln c_{+}(i, a)
\end{equation}
and
\[
-(x^++\frac{k}{2})\ln\frac{x^++\frac{k}{2}}{c_{-}(i, a)}+(\tilde x^++\frac{\tilde k}{2})\ln\frac{\tilde x^++\frac{\tilde k}{2}}{c_{-}(i,\tilde a)}\leq 
\Delta t\cdot\left|\frac{x^+-\tilde x^++\frac{1}{2}(k-\tilde k)}{\Delta t}\right|^{1-\alpha}+ 
\]
\begin{equation}\label{minusinequality}
+\big(\tilde x^++\frac{\tilde k}{2}\big)\cdot\ln\frac{c_{-}(i, a)}{c_{-}(i, \tilde a)}+\big(x^+-\tilde x^++\frac{1}{2}(k-\tilde k)\big)\cdot \ln c_{-}(i, a),
\end{equation}
where $\alpha \in (0,1)$.
Note that for notational simplicity, in some variables we removed the indices $i,j$
denoting dependence on the box.

For the second inequality of \eqref{ratioofnu} in all three cases we show that $|R_{i,j}^{\delta}(a)|<|R_{i,j}^{\delta}(\tilde a)|$.

\noindent
{\underline {\it Case 1:}\, \bf{The profile $a$ enters the safety zone.}} When the profile $a$ enters the safety zone, the new profile $\tilde a$ is defined as $\tilde a_{i,j-1}:=a_{i,j-1}$ and $\tilde a_{i,j}:=a_{i,j}-\delta'$. We choose $\tilde x^+:=x^+ $ and $\tilde k:=k-\frac{\delta'}{2}$, i.e., we keep the same number of plus jumps and we reduce the number of minus jumps. 
We also have that
\[
\tilde d=d-\frac{\delta' }{\Delta t}\,\,\,\,\,\mathrm{and}\,\,\,\,\,c_{\pm}(i,a)=c_{\pm}(i,\tilde a).
\]
So in \eqref{difference} 
there is no contribution to the error from the comparison of $x$ and $\tilde x$
and we only estimate the terms that
correspond to the number of minus, as in 
\eqref{minusinequality}.
Moreover, the last term in the r.h.s of \eqref{minusinequality} is negative.
Overall, we obtain
an upper bound for (\ref{difference}) given by
\begin{equation}\label{case1}
2\Delta t\left(\frac{\delta'}{2\Delta t}\right)^{1-\alpha}+\frac{\delta'}{2}.
\end{equation}

In addition, we have that $|R_{i,j}^{\delta}(a)|<|R_{i,j}^{\delta}(\tilde a)|$ since $m^{\delta}(\tilde a)=m^{\delta}(a)=0$ and $M^{\delta}_{i,j}(a)=k^{i,j-1}_{-}(a)-K^{i,j}-\delta\leq\tilde k^{i,j-1}_{-}(\tilde a)-\tilde K^{i,j}-\delta$.
Hence, by collecting the above estimates and substituting to \eqref{ratioofnu} we conclude that
\[
\frac{\nu^i_{{m_i((j-1)\Delta t)}}(
B^{\delta}_{i,j-1}(a))}{\nu^i_{{m_i((j-1)\Delta t)}}(
B^{\delta}_{i,j-1}(a))}\leq e^{\gamma^{-1}|I|\left(2\Delta t\left(\frac{\delta'}{2\Delta t}\right)^{1-\alpha}+\frac{\delta'}{2}\right)}.
\]
In this case, $M(\gamma)$ is given by the exponent in the right hand side.
As a general remark, we would like to stress that the above errors concern one space-time
box, so the overall error should be multiplied by the total number of boxes.
Furthermore, the changes in the given box influence all others as well and this has also
to be taken into account, but the error is similar as the one computed here. So 
we do not detail it here.

\bigskip

\noindent
{\underline {\it Case 2:}\, \bf{The profile $a$ exits the safety zone.}} Similarly to {\it Case 1}, 
the new profile is $\tilde a_{i,j-1}:=a_{i,j-1}-\delta'$ and $\tilde a_{i,j}:=a_{i,j}$. We choose $\tilde x^+=x^+-\frac{\delta'}{4}$ and $\tilde k:=k+\frac{\delta'}{2}>k$, i.e., we keep the same number of minus jumps and we decrease the number of plus jumps. Therefore, we have that
\[
\tilde d=d+\frac{\delta' }{\Delta t}\,\,\,\,\,\mathrm{and}\,\,\,\,\,|c_{\pm}(i,\tilde a)-c_{\pm}(i, a)|\leq \beta\delta' |I|,
\]
which implies that $|R_{i,j}^{\delta}(a)|\leq |R_{i,j}^{\delta}(\tilde a)|$ since $m^{\delta}(a)\geq 
m^{\delta}(\tilde a)$ and $k^{i,j-1}_{-}(a)-K^{i,j}$ is smaller or equal than all $k^{i,j-1}_{+}(a)$, $k^{i,j-1}_{+}(\tilde a)$ and $k^{i,j-1}_{-}(\tilde a)-\tilde K^{i,j}-\delta$. 
Hence, using inequalities \eqref{plusinequality} and \eqref{minusinequality} as also the rates have been altered
(in contrast to \underline{Case 1}),
we get the following upper bound for \eqref{difference}:
\[
\Delta t\left(\frac{\delta'}{4\Delta t}\right)^{1-a}+2\ln\left(1+\frac{\beta\delta' |I|}{c_m}\right)+2\beta|I|\delta'\Delta t+\frac{\delta'}{2}.
\]
Then, overall we have that 
\begin{eqnarray*}
\frac{\nu^i_{{m_i((j-1)\Delta t)}}(
B^{\delta}_{i,j-1}(a))}{\nu^i_{{m_i((j-1)\Delta t)}}(
B^{\delta}_{i,j-1}(a))}&\leq& e^{\gamma^{-1}|I|\left(\Delta t\left(\frac{\delta'}{4\Delta t}\right)^{1-a}+2\ln\left(1+\frac{\beta\delta' |I|}{c_m}\right)+2\beta|I|\delta'\Delta t+\frac{\delta'}{2}
\right)}.
\end{eqnarray*}

\bigskip

{\underline {\it Case 3:}\, \bf{Both $a_{i,j-1}$ and $a_{i,j}$ are in the safety zone.}} 
We subtract $\delta' $ from both $a_{i,j-1}$ and $a_{i,j}$, which also implies that $ \tilde d= d$. 
Hence, we choose
\[
\tilde x=x,\;\;\;\tilde k=k,
\]
which further implies that $|R_{i,j}^{\delta}(a)|\leq|R_{i,j}^{\delta}(\tilde a|)$ and $|c_{\pm}(i,\tilde a)-c_{\pm}(i, a)|\leq \beta\delta' |I|$. 
So the only terms in \eqref{difference}, \eqref{plusinequality} and \eqref{minusinequality} that contribute in the estimate are the terms which include the ratio and the difference of the rates. Thus, in this case, we obtain that:\begin{eqnarray*}
\frac{\nu^i_{{m_i((j-1)\Delta t)}}(
B^{\delta}_{i,j-1}(a))}{\nu^i_{{m_i((j-1)\Delta t)}}(
B^{\delta}_{i,j-1}(a))}&\leq&e^{\gamma^{-1}|I|\left(2\ln\left(1+\frac{\beta\delta' |I|}{c_m}\right)+2\beta|I|\delta'\Delta t\right)}.
\end{eqnarray*}
With this we conclude the proof of Lemma~\ref{movedprofile} as $\gamma\frac{\epsilon^{-3}}{|I|\Delta t}M(\gamma)\lesssim\epsilon^{-3}(\eta_3^{1-\alpha}+\eta_3 |I|)\to 0$ as $\gamma\to 0$.
\qed

\begin{remark1}
In some realizations and some boxes, it may also happen that the number of plus or minus jumps is finite. We show that in such a case we can still 
work with profiles away from $\pm 1$. Consider {\underline {Case 1}} with finite plus jumps 
when $a$ is close to $+1$. The other cases can be done similarly. Then, in \eqref{fullnu} for the probability of plus jumps
$P_{\gamma^{-1}|I| c_{+}(i,a)}(N_{i,j-1}^{-}=n_{-}^{i,j-1})$, as given in \eqref{occurrence},
we use the injective map $\iota$ as in \underline{Case 1} and obtain
\[
\frac{e^{-\gamma^{-1}|I|\bar c_{-}(i,a)\Delta t}}{e^{-\gamma^{-1}|I|\bar c_{-}(i,\tilde a)\Delta t}}\times\frac{(\gamma^{-1}|I|\bar c_{-}(i,a)\Delta t )^{(n^{+}_{i,j-1}+\frac{\gamma^{-1}|I|k^{i,j}}{2})}}{(\gamma^{-1}|I|\bar c_{-}(i,\tilde a)\Delta t )^{(\tilde n^{+}_{i,j-1}+\frac{\gamma^{-1}|I|\tilde k^{i,j}}{2})}}\times\frac{(\tilde n^{+}_{i,j-1}+\frac{\gamma^{-1}|I|\tilde k^{i,j}}{2})!}{(n^{+}_{i,j-1}+\frac{\gamma^{-1}|I|k^{i,j}}{2})!}\leq
\]
\[
\leq \left(\gamma^{-1}|I|\Delta t\times \gamma^{-1}|I|\right)^{\gamma^{-1}|I|\frac{\delta'}{4}},
\]
because the rates for $a$ and $\tilde a$ are equal for the {\underline {Case 1}}. Taking the logarithm of this error multiplied by the number of coarse-grained boxes, $\epsilon^{-3}/|I|\Delta t$,
and multiplying by $\gamma$ we get a vanishing number as $\gamma\to 0$:
\[
\gamma\frac{\epsilon^{-3}}{|I|\Delta t}\gamma^{-1}|I|\frac{\delta'}{4}\ln\left(\gamma^{-1}|I|\Delta t\times \gamma^{-1}|I|\right),
\]
since  $\delta'=\Delta t\cdot\eta_3$ and $\eta_3\cdot\epsilon^{-3}\to 0$. 
\end{remark1}

\bigskip
\bigskip

\end{document}